\documentclass[aoas,preprint]{format}

\usepackage{amsthm,amsmath,natbib}
\usepackage[colorlinks,citecolor=blue,urlcolor=blue]{hyperref}
\usepackage{amsfonts,amssymb}

\newtheorem{Algorithm}{Algorithm}
\usepackage{graphicx}
\usepackage[font={small,it}]{caption}
\usepackage{array}
\usepackage{soul}
\usepackage{pdfpages}
\startlocaldefs
\usepackage{relsize}
\usepackage{geometry}
\usepackage{fge}
\usepackage{enumitem}
\usepackage{pdfpages}
\newcommand\numberthis{\addtocounter{equation}{1}\tag{\theequation}}
\endlocaldefs

\begin{document}

\begin{frontmatter}

\title{Two-way sparsity for time-varying networks, with applications in genomics}
\runtitle{Two-way sparsity, for time-varying networks}

\begin{aug}
\author{\fnms{Thomas E.} \snm{Bartlett}\thanksref{t1,m1}\ead[label=e1]{thomas.bartlett.10@ucl.ac.uk}},
\and
\author{\fnms{Ioannis} \snm{Kosmidis}\thanksref{m2,m3}}
\and
\author{\fnms{Ricardo} \snm{Silva}\thanksref{m1,m3}}

\thankstext{t1}{thomas.bartlett.10@ucl.ac.uk}
\runauthor{T.E. Bartlett et al}
\affiliation{Department of Statistics, University College London, WC1E 6BT, UK\thanksmark{m1}}
\affiliation{Department of Statistics, University of Warwick, Coventry, CV4 7AL, UK\thanksmark{m2}}
\affiliation{The Alan Turing Institute, London, NW1 2DB, UK\thanksmark{m3}}

\end{aug}

\begin{abstract}
We propose a novel way of modelling time-varying networks, by inducing two-way sparsity on local models of node connectivity. This two-way sparsity separately promotes sparsity across time and sparsity across variables (within time). Separation of these two types of sparsity is achieved through a novel prior structure, which draws on ideas from the Bayesian lasso and from copula modelling. We provide an efficient implementation of the proposed model via a Gibbs sampler, and we apply the model to data from neural development. In doing so, we demonstrate that the proposed model is able to identify changes in genomic network structure that match current biological knowledge. Such changes in genomic network structure can then be used by neuro-biologists to identify potential targets for further experimental investigation.
\end{abstract}


\end{frontmatter}

\section{Introduction}

\begin{figure}[t]
\vspace{-0ex}
\centering\includegraphics[width=0.95\textwidth]{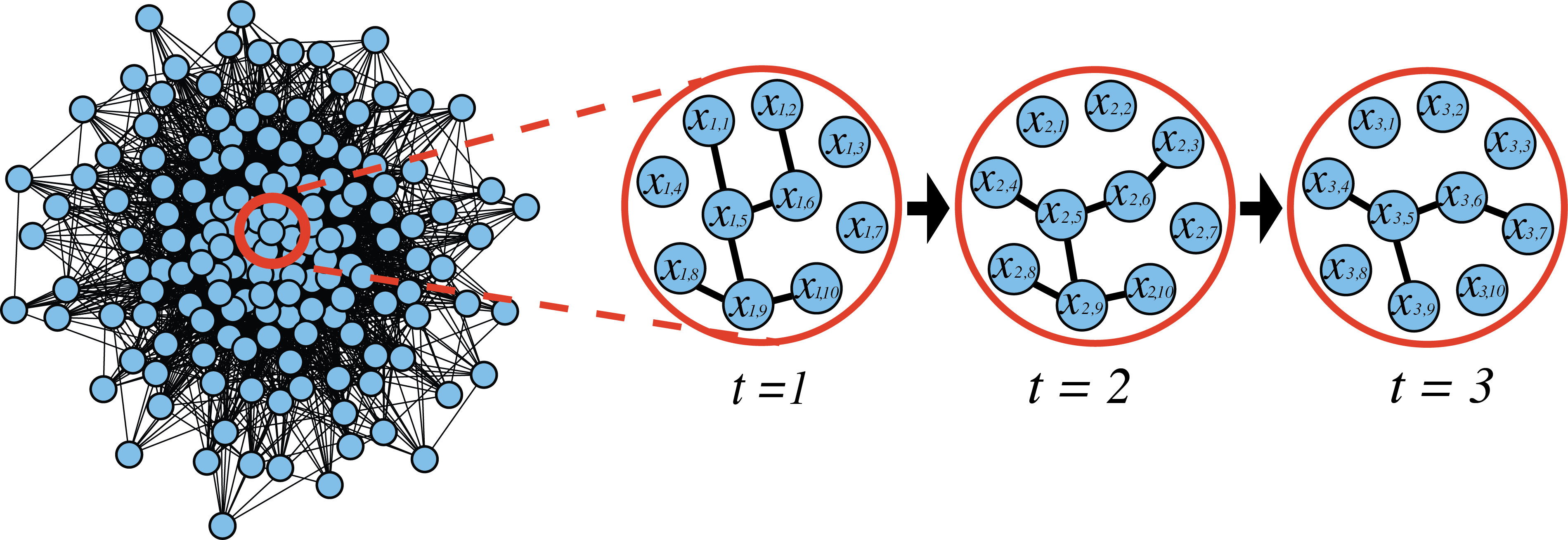}
\vspace{-1ex}
\caption{Model of time-varying network structure. Each $x_{t,i}$ represents a class label or continuous variable for node $i$ (e.g., the expression-level of gene $i$) at time $t$. The links represent network interactions or dependencies between $x_{1,1}, x_{1,2},...$ (e.g. due to gene regulation), which may be different to those between $x_{2,1}, x_{2,2},...$ and $x_{3,1}, x_{3,2},...$. Hence, these network interactions may vary with time.}\label{dynNetFig}
\vspace{-1ex}
\end{figure}

Network models have become an important topic in modern statistics, and the evolution of network structure over time (illustrated in Figure \ref{dynNetFig}) is an important area of study. Network structures that evolve over time naturally occur in a range of applications. Examples of recent applications include evolving patterns of human interaction \citep{durante2016locally} such as in social networks \citep{sekara2016fundamental}, time-varying patterns of interaction between genes and their protein-products in biological networks \citep{alexander2009understanding,lebre2010statistical}, and time-varying patterns of connectivity in the brain \citep{schaefer2014dynamic}. However, network models with temporal structure have only recently begun to be studied in detail in statistical research.

An important application area of statistical network models is genomics. Network models are a natural way to describe and analyse patterns of interactions (represented by network edges) between genes and their protein-products (represented by network nodes). An important interaction of this type is gene regulation, in which the protein-product of one gene influences the output level of the protein-product of another gene. Much gene regulation is characteristic of a particular cell type, so that a cell knows its role within the organism. These specific regulatory network structures that are characteristic of particular cell-types are established during embryonic development. Changes in normal gene regulation are also inherent to cancer progression, so that cells \textquoteleft forget\textquoteright\ how they should act, taking on pathological roles (regulatory network re-wiring) \citep{suva2014reconstructing}. However, whilst network models are well established in genomics, historically these models have typically been static, ignoring the fact that genomic processes are inherently time-varying.

There are many examples of recent work on models of time-varying networks. In statistics, this work covers methods based on Markov processes \citep{crane2016dynamic}, on dynamic Erd\H{o}s-R\'{e}nyi graphs \citep{rosengren2016dynamic}, and on sparse regression methods \citep{kolar2010estimating}. It also includes work on time-varying community structure \citep{zhang2012common}, on methods extending the stochastic block model \citep{xu2013dynamic,matias2016statistical}, and related non-parametric graphon-based methods \citep{pensky2016dynamic}, as well as non-parametric methods for dynamic link prediction \citep{sarkar2014nonparametric} and methods from Bayesian nonparametrics \citep{palla2016bayesian}. Other related work includes sparse graphical models that can take account of different time-points \citep{kalaitzis2013bigraphical}.

Motivated by genomics applications, we propose a novel framework for modelling time-varying networks, by inducing \emph{two-way sparsity} on local models of the connectivity of each node to all the others. This is achieved as follows. We start with a regression likelihood function that assumes that observations are mutually independent over time. Dependence is then induced through a novel prior structure that promotes sparsity in a two ways: across time, and within time. This decoupling of the induced sparsity is achieved through a copula specification for the parameters in the likelihood function. Specifically, the regression coefficients for one node across different time-points are jointly distributed according to a Gaussian copula with Laplace marginal distributions. The correlation matrix of the Gaussian copula is formed by assuming that the correlation between time-points decays with time in a structured, parsimonious way that also ensures its positive definiteness. In this correlation matrix, the only free parameter is the correlation between consecutive time-points, which is given a reverse-exponential prior distribution with support in $[0, 1)$. This prior on the correlation across time discourages large differences in the regression coefficients between consecutive time-points and, as a consequence, also discourages large changes in the inferred structure of the network. 

The decoupling of the marginal and dependence structure that is facilitated by the copula specification, and the particular form of the correlation matrix, allow for precise control of marginal priors. This decoupling also makes the adoption of generalisations of the Bayesian lasso, such as the horseshoe \citep{carvalho2010horseshoe}, easy to implement in place of the marginal Bayesian lasso prior that we use. The prior dependence among parameters across time can also be viewed as a Bayesian version of the fused lasso \citep{tibshirani2005sparsity}, while within each time-slice we directly utilise existing work on the Bayesian lasso \citep{park2008bayesian}. In fact, the proposed modelling framework has the Bayesian lasso as a special case, when the correlation between time-points is set to zero. From a frequentist point of view, the sparsity structure we propose would fall within the remit of the generalised lasso \citep{tibshirani2011solution}, which has the fused lasso as a special case \citep{tibshirani2005sparsity}. Bayesian versions of the fused lasso have also been proposed \citep{kyung2010penalized,shimamura2016bayesian}. However, a key difference between those methods and the modelling framework we propose, is the formal decoupling of sparsity across time (which the fused lasso induces), from sparsity within time. Importantly, we are able to apply this proposed modelling framework locally to each network node, as previous authors have done \citep{kolar2010estimating}. Because these local model fits are mutually independent they can easily be carried out sequentially or in parallel, meaning that in practice, we are able to work with large networks of tens of thousands of nodes. The novel prior structure proposed, which enables the time-varying network inference, is also of interest more generally beyond the context of network science. This novel prior structure is relevant in any context where sparse regression with time-varying regression parameters is desirable.

The rest of the paper is structured as follows. In Section \ref{methodsSec}, we set up notation, and specify the model. Then, in Section \ref{simSec} we present the results of fitting the model to simulated data, and in Section \ref{resSec} we present the results of fitting the model to single-cell transcriptome data. Finally, in Section \ref{discussSect}, we summarise our findings and discuss their broader context. The Supplementary Material we provide all proofs and derivations, data pre-processing details, and Supplementary Figures, as well as a freely available software implementation of our proposed model and algorithm.

\section{Proposed methodology}\label{methodsSec}
\subsection{Data description}\label{dataDescripSec}
The two-way sparsity that is induced by the proposed modelling framework is motivated by the problem of inferring time-varying structure in genomic networks. In these networks nodes represent genes: for each node there are observations or measurements of the activity level of the corresponding gene (the `gene-expression level'). These node-specific observations make up the data-set. The expression-level of a gene is generally influenced by the expression-level of several other genes (in a process called `gene regulation'). Hence, a natural application for models of time-varying networks is understanding dynamic patterns of gene-regulation in biological processes, such as neural development. Measurements of gene transcript counts are often used as a surrogate for gene expression level in RNA sequencing data, and hence we base our real-data example on single-cell transcriptomic data. We note that single-cell transcriptomic data is a type of single-cell gene-expression data.

Single-cell gene-expression data are ideal for this application, because data from a study of this type will typically be obtained from a heterogeneous mixture of cells, each of which may be at a different point on a trajectory through the biological process under investigation. For example, in the context of neural development, some of these cells may be stem-cells, whereas some may be fully differentiated cells (e.g., neurons), with a whole spectrum of cells in between. Each cell can be thought of as an independent sample from the underlying latent biological process; in this example, that process is neural development. Thus, we can think of the progression of a cell through this process of neural development in terms of a `developmental trajectory'. The progression along such a developmental trajectory can be quantified in terms of `developmental time', which is simply a measure of a temporally-ordered progression through the process of cellular development. For each of the cell-samples in the data, no information is available other than its high-dimensional gene-expression measurements. Hence, it is necessary to first infer the `developmental time' of the cell-samples before fitting any time-varying network model. This leads to an ordered sequence of pseudo-temporal measurements $x_{1,i}, x_{2,i},...x_{t,i},...,x_{T,i}$ of the log-expression level for gene $i$. Importantly, the $x_{1,i}, x_{2,i},...\text{etc}$ are taken from different cell samples for each pseudo-time point, and are hence independent. Inference like this is more generally referred to as `pseudo-time' inference, and several methods exist to carry it out: see for example work by \cite{qiu2011extracting} and \cite{trapnell2014dynamics}.

\subsection{Model overview}\label{modelOverviewSec}
We develop a model for each target-node conditional on all the other nodes, and then we apply this model to several target-nodes of interest. This is different from, for example, the work of \cite{friedman2008sparse} and \cite{fan2009network} who consider the modelling of all nodes jointly. Such a target-node approach has been used previously by \cite{kolar2010estimating}, and it allows the network structure to be inferred independently around each target-node $i\in\{1,...,p\}$. This strategy has several advantages. Firstly, variable screening can be applied before model fitting. This allows the dimensionality of the problem to be reduced from $p$ of the order of tens of thousands down to $p'$ of the order of a few hundred for each parallel model fit around a target-node, whilst still allowing the global network structure to be estimated over tens of thousands of target-nodes, if required. Our modelling strategy also allows the local network structure to be estimated around only a small number of target-nodes if required, controlling computational expense, whilst still inferring the connected node-sets from tens of thousands of nodes. Inference is carried out with a sparse linear model, taking the observations for node $i$ at time $t$ as the response, and the observations for all nodes $j\neq i$ at time $t$ as potential predictors. From these potential predictors, the set of predictors `chosen' by the sparse model fit are then used to infer the network structure. Specifically, we want to infer the network structure around a fixed set of nodes with a set of edges that varies with time. In this scenario, only the patterns of interconnectivity change as the network evolves (Figure \ref{dynNetFig}), which is the scenario most relevant to genomics applications. Such a network can be represented with a time-varying adjacency matrix $\mathbf{A}$, where $A_{i,j,t}$ denotes the absence ($A_{i,j,t}=0$) or presence ($A_{i,j,t}>0$) of an edge between nodes $i$ and $j$ at time $t$. We note that under this scheme, the local model fit (which is responsible for the computational load) does not depend on the network estimation (which takes place subsequently). The inferred network is a particular summary of the posteriors that are obtained from several of our model fits. We propose a model for node-wise regression, and we suggest how to summarise these models over several nodes of a network.

\subsection{Model likelihood}\label{mainModel}
We assume a likelihood function where observations are mutually independent over time. This is an assumption that is compatible with high-dimensional gene-expression data, where no single cell can be measured at more than one time-point. We note that this implies that observations are independent at different time-points. Let $\mathbf{X}$ represent the full data-set for the nodes shown in Figure \ref{dynNetFig}, with time varying down the rows, and with each node corresponding to a different column. Then, $x_{t,i}$ denotes the value for some node in the system at time $t\in\{1,...,T\}$, for $i\in\{1,...,p\}$, and the row-vector $\mathbf{x}_{t,\fgebackslash i}$ denotes the values for the other $p-1$ nodes at time $t$. We model the dependence of $x_{t,i}$ on $\mathbf{x}_{t,\fgebackslash i}$ as:
\begin{equation}
x_{t,i}=a_i+\mathbf{b}^{(i)}_{t,:}\ \mathbf{x}_{t,\fgebackslash i}^\top+\epsilon_{t,i},\label{mainLinModelEq}
\end{equation} 
where $\mathbf{b}^{(i)}_{t,:}$ is a vector of linear model parameters, and $\epsilon_{t,i}\sim\mathcal{N}\left(0,\tau_i^{-1}\right)$. 

The response variable $x_{t,i}$ corresponds to the observations for a `target' node around which we are modelling the local network structure, whereas the variables represented by $\mathbf{x}_{t,\fgebackslash i}$ correspond to the observations for all the other nodes of the network. To model the whole network, we must fit model (\ref{mainLinModelEq}) around each target-node in turn. We note that here we make an assumption about the existence of a global undirected Markov network \citep{lauritzen1996graphical} that explains the independence constraints in the model. This assumption has also been used previously by \cite{kolar2010estimating} in an equivalent context. We note that our approach does not enforce hard constraints, such as $\mathbf{b}^{(i)}_{t,j}=\mathbf{b}^{(j)}_{t,i}$. However, it is computationally very expensive to work with a global, coherent model, where such constraints can be enforced. In this work, we have opted to sacrifice some coherence for the sake of computational efficiency. This enables us to estimate quantities of interest in a computationally-efficient manner through an overparameterized representation of a joint model. It also enables us to focus on a particular subset of nodes of interest without having to go through an overly-expensive computation for the estimation of a global, coherent model.

Using $\mathbf{b}^{(i)}_{:,j}$ to denote the column-vector of model parameters for covariate $j$ for $t\in\{1,...,T\}$, we collect parameters in matrix $\mathbf{B}^{(i)}=\left[\mathbf{b}^{(i)}_{:,1},\mathbf{b}^{(i)}_{:,2},...,\mathbf{b}^{(i)}_{:,p-1}\right]$. In the next section, we postulate a prior for dependencies within each column $j$ of $\mathbf{B}^{(i)}$, whilst noting that the columns of $\mathbf{B}^{(i)}$ (each corresponding to a different node as covariate) are independent of each other. We also introduce the notation $x_{t,i,k}$ and $\mathbf{x}_{t,\fgebackslash i,k}$ to represent observations of $x_{t,i}$ and $\mathbf{x}_{t,\fgebackslash i}$ for sample $k\in\{1,...,n_t\}$ at time $t$.

We denote {$\mathbf{x}_{:,i}=[x_{1,i,1},...,x_{1,i,n_1},...,x_{t,i,1},...,x_{t,i,n_t},...,x_{T,i,1},...,x_{T,i,n_T}]^\top$} and\\ {$\mathbf{X}_{:,\fgebackslash i}=[\mathbf{x}_{1,\fgebackslash i,1}^\top,\mathbf{x}_{1,\fgebackslash i,n_1}^\top,...,\mathbf{x}_{t,\fgebackslash i,1}^\top,...,\mathbf{x}_{t,\fgebackslash i,n_t}^\top,...,\mathbf{x}_{T,\fgebackslash i,1}^\top,...,\mathbf{x}_{T,\fgebackslash i,n_T}^\top]^\top$}, where $\mathbf{x}_{:,i}$ is column $i$ of data-matrix $\mathbf{X}$, and $\mathbf{X}_{:,\fgebackslash i}$ is data-matrix $\mathbf{X}$ without column $i$. Hence, we can write the model likelihood for the target-node $i$ as:
\begin{equation*}
P(\mathbf{x}_{:,i}|\mathbf{X}_{:,\fgebackslash i},\mathbf{B}^{(i)},a_i,\tau_i)=\prod_{t=1}^T\prod_{k=1}^{n_t}\sqrt{\frac{\tau_i}{2\pi}}e^{-\tau_i(x_{t,i,k}-\mathbf{b}^{(i)}_{t,:}\cdot\mathbf{x}_{t,\fgebackslash i,k}^\top-a_i)^2/2}.\numberthis\label{dataCondParams}
\end{equation*}
We note that we consider likelihoods in the form of equation (\ref{dataCondParams}) for each target-node $i$. 

\subsection{Priors with decoupled two-way sparsity}\label{priorDef}
We model the regression coefficients $\mathbf{b}^{(i)}_{:,j}$ across time-points $t=1,...,T$ with a Gaussian copula with Laplace marginal distributions, as follows. The elements of $\mathbf{b}^{(i)}_{t,:}$ $(t = 1,...,T)$ are marginally distributed as $b^{(i)}_{t,j}\sim\text{Laplace }(1/\lambda)$, with probability density function $\frac{\lambda}{2}e^{-\lambda|\cdot|}$ and cumulative distribution function $F_\mathcal{L}[b^{(i)}_{t,j}]$, for $t\in\{1,...,T\}$ and $j\in\{1,...,p-1\}$. Hence,
$\Phi^{-1}\left\{F_\mathcal{L}[b^{(i)}_{t,j}]\right\}$ follows a Gaussian distribution for $t\in\{1,...,T\}$ and $j\in\{1,...,p-1\}$, where $\Phi$ is the standard-normal cumulative distribution function. The dependencies between $\Phi^{-1}\left\{F_\mathcal{L}[b^{(i)}_{t,j}]\right\}$ and $\Phi^{-1}\left\{F_\mathcal{L}[b^{(i)}_{t+1,j}]\right\}$ are then modelled through their joint distribution as:
\begin{equation*}
{\footnotesize
\left[\begin{array}{c}\vspace{1ex} \Phi^{-1}\left\{F_\mathcal{L}[b^{(i)}_{1,j}]\right\} \\\vspace{1ex}\Phi^{-1}\left\{F_\mathcal{L}[b^{(i)}_{2,j}]\right\} \\\vspace{1ex} \vdots \\\vspace{1ex} \Phi^{-1}\left\{F_\mathcal{L}[b^{(i)}_{T,j}]\right\} \end{array} \right]\sim\mathcal{N}\left(\mathbf{0},\boldsymbol{\Sigma}^{(i)}_j\right),\enspace\text{with}\enspace\boldsymbol{\Sigma}^{(i)}_j=\left[\begin{array}{ccccc}\vspace{1.3ex}1&\rho^{(i)}_j&\left(\rho^{(i)}_j\right)^2&\cdots&\left(\rho^{(i)}_j\right)^T\\\vspace{1.3ex}\rho^{(i)}_j&1&\rho^{(i)}_j&\cdots&\left(\rho^{(i)}_j\right)^{T-1}\\\vspace{1.3ex}\left(\rho^{(i)}_j\right)^2&\rho^{(i)}_j&1&\cdots&\left(\rho^{(i)}_j\right)^{T-2}\\\vspace{1.3ex}\vdots&\vdots&\vdots&\ddots&\vdots\\\vspace{1.3ex}\left(\rho^{(i)}_j\right)^T&\left(\rho^{(i)}_j\right)^{T-1}&\left(\rho^{(i)}_j\right)^{T-2}&\cdots&1\end{array}\right],\numberthis\label{fullCovSpec}
}
\end{equation*}
and hence the regression coefficients are modelled as a Gaussian copula:
\begin{equation*}
F\left[\mathbf{b}^{(i)}_{:,j}\right]=\Phi_p\left[\Phi^{-1}\left\{F_\mathcal{L}[b^{(i)}_{1,j}]\right\},\Phi^{-1}\left\{F_\mathcal{L}[b^{(i)}_{2,j}]\right\},...,\Phi^{-1}\left\{F_\mathcal{L}[b^{(i)}_{T,j}]\right\};\boldsymbol{\Sigma}^{(i)}_j\right].
\end{equation*}
The correlation parameter $\rho^{(i)}_j$ is assumed to have a reverse-exponential distribution with support $[0,1)$ and density
\begin{equation}
f_\text{rexp}[\rho^{(i)}_j]\sim ke^{k\rho^{(i)}_j}/(e^k-1).\label{revExpPriorDef}
\end{equation}

\begin{figure}[t]
\vspace{0ex}
\centering\includegraphics[width=0.5\textwidth]{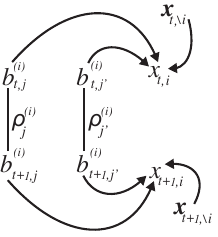}
\vspace{-1ex}
\caption{Chain graphical model \citep{lauritzen1996graphical}. The diagram shows the dependence of $x_{t,i}$ (the value of the target-node at time $t$) on $\mathbf{x}_{t,\fgebackslash i}$ (which represents the values of two other nodes $j$ and $j'$ at time $t$), and on the corresponding model parameters $b^{(i)}_{t,j}$ and $b^{(i)}_{t,j'}$. Model parameters are correlated across time, such that $\Phi^{-1}\left\{F_\mathcal{L}[b^{(i)}_{t,j}]\right\}$ and $\Phi^{-1}\left\{F_\mathcal{L}[b^{(i)}_{t+1,j}]\right\}$ have correlation $\rho^{(i)}_j$.}\label{graphicalModel}
\vspace{-1ex}
\end{figure}	

The structure of $\boldsymbol{\Sigma}^{(i)}_j$ is such that transformed model parameters at adjacent points in time, such as $\Phi^{-1}\left\{F_\mathcal{L}[b^{(i)}_{t,j}]\right\}$ and $\Phi^{-1}\left\{F_\mathcal{L}[b^{(i)}_{t+1,j}]\right\}$, have correlation $\rho^{(i)}_j$ (Figure \ref{graphicalModel}). Then, the transformed parameters separated by two time-points have correlation $(\rho^{(i)}_j)^2$, etc. Thus, also denoting the sequence of transformed model parameters $\Phi^{-1}\left\{F_\mathcal{L}[b^{(i)}_{1,j}]\right\}$, $\Phi^{-1}\left\{F_\mathcal{L}[b^{(i)}_{2,j}]\right\}$,...,$\Phi^{-1}\left\{F_\mathcal{L}[b^{(i)}_{t,j}]\right\}$ forms a Markov chain, meaning that $\boldsymbol{\Sigma}^{(i)}_j$ is guaranteed to be positive-definite for $\rho^{(i)}_j\in[0,1)$, and by construction
\begin{equation}
b^{(i)}_{t+1,j}\perp b^{(i)}_{t-1,j},b^{(i)}_{t-2,j},...|b^{(i)}_{t,j}.\label{thetaCondIndepState}
\end{equation}
\begin{figure}[t!]
\vspace{0ex}
\centering\includegraphics[width=0.75\textwidth]{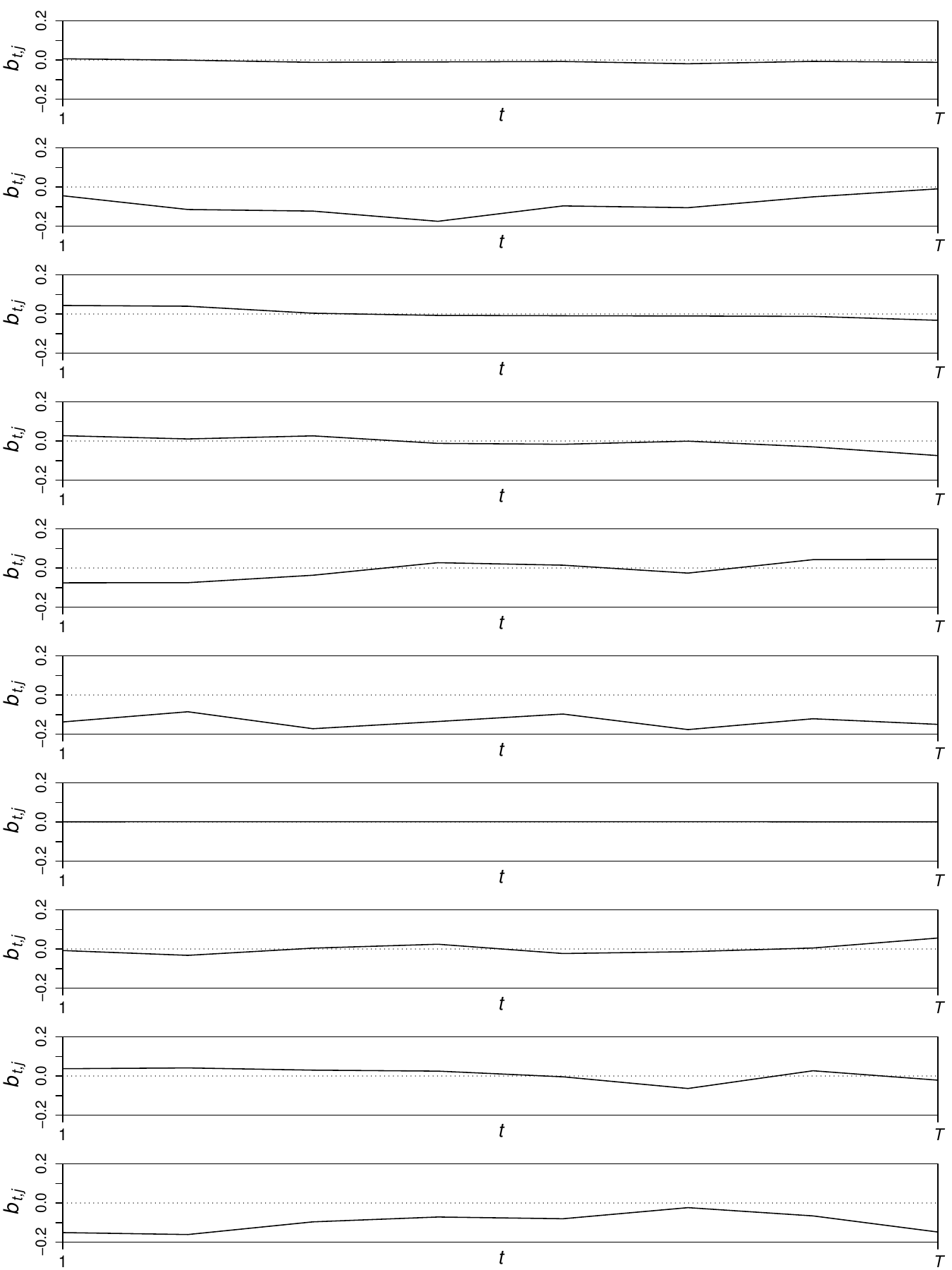}
\vspace{-1ex}
\caption{Samples from the prior on $\mathbf{b}_{t,j}$ plotted against $t$, illustrating their correlation structure over time. These results are with $T=8$, $\lambda=20$ and $k=1$.}\label{priorTrajSamp}
\vspace{-1ex}
\end{figure}
\noindent
Such a construction for $\boldsymbol{\Sigma}^{(i)}_j$ discourages differences in the regression coefficients for the same covariate between adjacent time-points, and hence also discourages changes in the network structure over time, resulting in sparsity across time. Then, transforming the $\Phi^{-1}\left\{F_\mathcal{L}[b^{(i)}_{t,j}]\right\}$ back to $b^{(i)}_{t,j}$, where the $b^{(i)}_{t,j}$ are marginally Laplace distributed, achieves sparsity within time by discouraging regression coefficients from taking non-zero values, hence also encouraging discovery of sparse network structures. Figure \ref{priorTrajSamp} shows ten samples from our proposed prior on $b^{(i)}_{t,j}$ plotted against $t$, and demonstrates the correlation structure enforced by the prior over time. We again note that while $b^{(i)}_{t,j}$ and $b^{(i)}_{t+1,j}$ are correlated, $b^{(i)}_{t,j}$ and $b^{(i)}_{t,j'}$ are independent.

Recent work by \cite{shimamura2016bayesian} that takes a Bayesian approach to generalising the fused lasso could be used similarly to the approach we propose, by modelling the same set of covariates at multiple time-points whilst enforcing smooth changes across time as well as sparsity overall. However, \cite{shimamura2016bayesian} achieve their result by simply multiplying together separate frequentist-inspired priors for smoothness across time and for sparsity. Specifically, they multiply together a Laplace prior to penalise individual non-zero model parameters, with the ultra-sparse negative-exponential-gamma (NEG) prior to penalise non-zero differences in parameters. The Laplace-NEG prior is defined (choosing notation to be consistent with that of our model) as:
\begin{equation}
P(\mathbf{b}^{(i)}_{:,j})\propto\prod_{t=1}^T\text{Laplace}(b^{(i)}_{t,j}|\lambda)\prod_{t=2}^T\text{NEG}(b^{(i)}_{t,j}-b^{(i)}_{t-1,j}|\lambda^\dagger,\gamma)\label{laplNegDist},
\end{equation}
where the Laplace density is defined as $\frac{\lambda}{2}e^{-\lambda|\cdot|}$, and
\begin{equation*}
\text{NEG}(\cdot|\lambda^\dagger,\gamma)=\int_0^\infty\int_0^\infty f_\mathcal{N}(\cdot|0,\tau^2)f_\mathcal{\gamma}(\tau^2|1,1/\psi)f_\mathcal{\gamma}(\psi|\lambda^\dagger,1/\gamma^2)d\tau^2d\psi,
\end{equation*}
where $ f_\mathcal{N}$ and $f_\mathcal{\gamma}$ are the Normal and Gamma densities, respectively. Sampling from the distribution of equation (\ref{laplNegDist}) is done by simulating exponential and gamma random variables, which are then used to form the precision matrix of a multivariate normal distribution, as specified by \cite{shimamura2016bayesian}. In contrast to the Laplace-NEG prior, the model we propose retains the property that, marginally, each parameter still follows the Bayesian lasso prior \citep{park2008bayesian}. In particular, if we set $\rho^{(i)}_j=0$ (for $j=1,2,...,p-1$), then the model we propose is exactly the same as the Bayesian lasso. This is important because it makes it easier to set priors, including variants of the Bayesian lasso that avoid its well-known shortcomings (see for example the work by \cite{castillo2015bayesian} and \cite{van2016conditions}). Although we will not consider such variants here, they follow directly by mimicking the construction using the Bayesian lasso.

\begin{figure}[t]
\vspace{0ex}
\centering\includegraphics[width=0.95\textwidth]{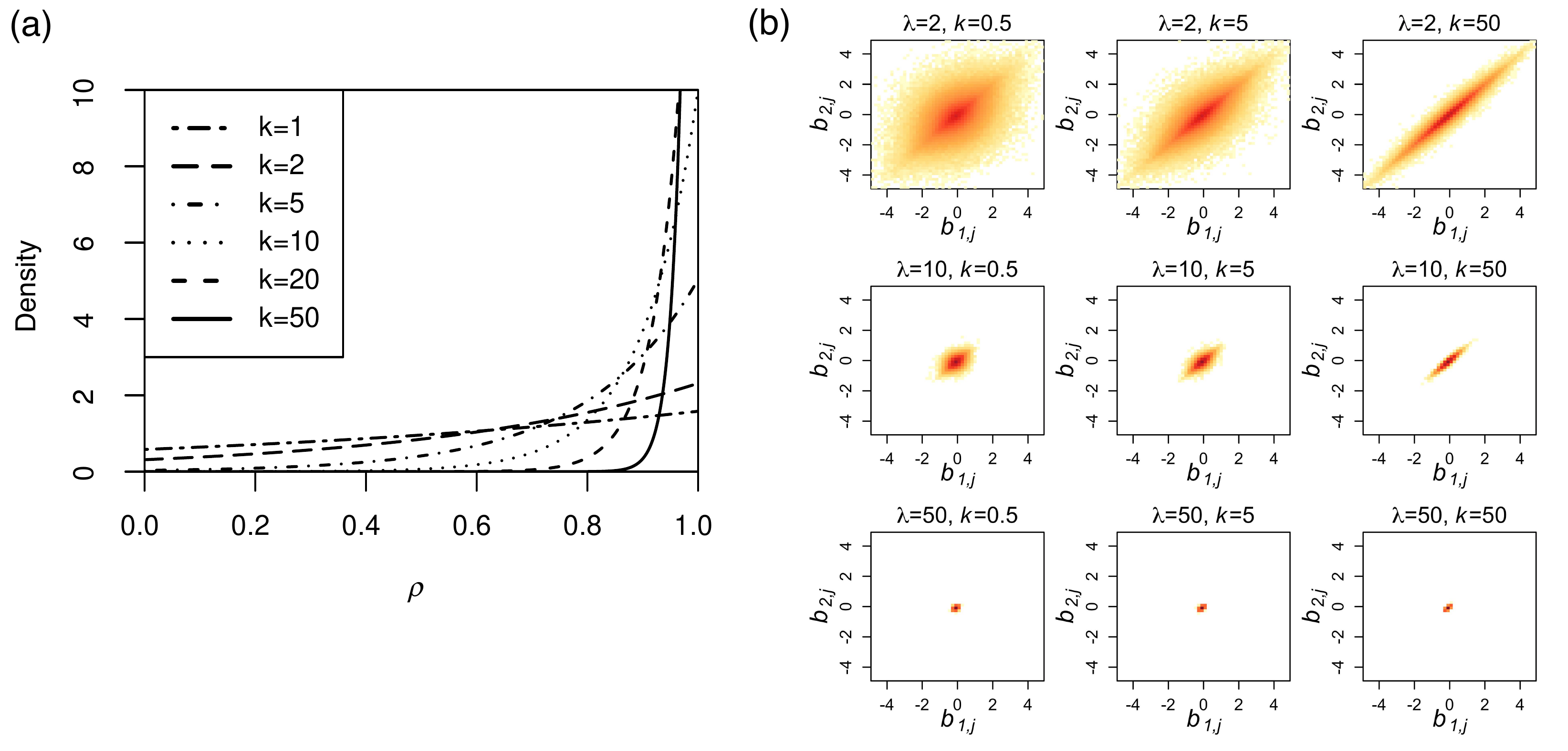}
\vspace{-1ex}
\caption{(a) Density function of the reverse-exponential prior. (b) Heatmaps of the bivariate log-densities of prior samples for $\mathbf{b}^{(i)}_{:,j}=[b^{(i)}_{1,j},b^{(i)}_{2,j}]^\top$.}\label{priorFig}
\vspace{-1ex}
\end{figure}

The novel prior we use on $\rho^{(i)}_j$ is a `reverse exponential prior' (equation (\ref{revExpPriorDef})). Figure \ref{priorFig}a shows the probability density function of the reverse-exponential prior for different values of hyper-parameter $k$. Figure \ref{priorFig}b then shows heatmaps of the bivariate density distributions of samples from the decoupled-sparsity prior for a parameter $j$ over two time-points, i.e., $\mathbf{b}^{(i)}_{:,j}=[b^{(i)}_{1,j},b^{(i)}_{2,j}]^\top$, for a range of values of $\lambda$ and $k$ (the corresponding marginal densities are shown in Figures S6 and S7 in the Supplementary Information). For comparison, Figures S3 - S5 in Supplement D show samples from the Laplace-NEG prior as defined in equation (\ref{laplNegDist}), for various values of $\lambda$ (which acts equivalently to $\lambda$ in our model, controlling sparsity of individual model parameters), and various values of $\lambda^\dagger$ and $\gamma$ (which both act equivalently to $k$ in our model, controlling sparsity of differences between model parameters). The main difference between these priors is that our decoupled-sparsity prior still marginally follows the Bayesian lasso prior, and is hence a direct generalisation of the Bayesian lasso to this setting with time-varying model parameters. In other words, our prior does what the Laplace-NEG does, but with the added benefit that we generalise the Bayesian lasso.

\subsection{Posterior inference}\label{postInf}
The order-1 Markovian relations specified by equation (\ref{thetaCondIndepState}) are also computationally attractive, because they result in models with banded precision matrices. From equation (\ref{thetaCondIndepState}), and denoting $\theta^{(i)}_{t,j}=\Phi^{-1}\left\{F_\mathcal{L}[b^{(i)}_{t,j}]\right\}$ it follows that the partial correlation of $\theta^{(i)}_{t+m,j}$ with $\theta^{(i)}_{t+l,j}$ will be zero for all $|m-l|>1$. Hence, all entries of the precision matrix $[\boldsymbol{\Sigma}^{(i)}_j]^{-1}$ will be zero except the diagonal and the elements immediately adjacent to it (i.e., the sub- and super-diagonals). These relationships allow all the entries of this precision matrix to be found easily in terms of $\rho^{(i)}_j$ by solving $[\boldsymbol{\Sigma}^{(i)}_j]^{-1}\boldsymbol{\Sigma}^{(i)}_j=\mathbb{I}$, which gives:
\begin{equation}
\left(\left[\boldsymbol{\Sigma}^{(i)}_j\right]^{-1}\right)_{t,t^\prime}=
\begin{cases}
\vspace*{1ex}
 1/(1-[\rho^{(i)}_j]^2), & \text{if}\ t^\prime=t=1\ \text{or}\ t^\prime=t=T, \\
\vspace*{1ex}
 (1+[\rho^{(i)}_j]^2)/(1-[\rho^{(i)}_j]^2), & \text{if}\ t^\prime=t>1\ \text{and}\ t^\prime=t<T, \\
\vspace{1ex}
 -\rho^{(i)}_j/(1-[\rho^{(i)}_j]^2), & \text{if}\ t^\prime=t+1\ \text{or}\ t^\prime=t-1, \\
 0, & \text{otherwise}\label{precMatSpec}
\end{cases}
\end{equation}
where $\left(\left[\boldsymbol{\Sigma}^{(i)}_j\right]^{-1}\right)_{t,t^\prime}$ represents the $(t,t^\prime)$ element of the precision matrix $[\boldsymbol{\Sigma}^{(i)}_j]^{-1}$. A full derivation of equation (\ref{precMatSpec}) is given in Supplement A.

The model parameters $\mathbf{B}^{(i)}$ can be sampled directly from multivariate Normal distributions, without needing the intermediate transformation to the marginally Laplace-distributed variables described in Section \ref{priorDef}. This can be achieved with an algebraic manipulation which is an extension from the Bayesian lasso, as follows. The Laplace distribution can be written as an uncountable mixture of zero-mean Normal distributions, with the variances of the mixture components distributed as $\text{Exp}(\frac{\lambda^2}{2})$ \citep{andrews1974scale,park2008bayesian}. Specifically,
\begin{equation*}
P(b^{(i)}_{t,j}|\lambda)=\frac{\lambda}{2}e^{-\lambda|b^{(i)}_{t,j}|}=\int_0^\infty P(b^{(i)}_{t,j},s^{(i)}_j|\lambda)ds^{(i)}_j,
\end{equation*}
where
\begin{equation*}
P(b^{(i)}_{t,j},s^{(i)}_j|\lambda)=\frac{1}{\sqrt{2\pi s^{(i)}_j}}e^{-[b^{(i)}_{t,j}]^2/[2s^{(i)}_j]}\frac{\lambda^2}{2}e^{-\lambda^2s^{(i)}_j/2},
\end{equation*}
for $s^{(i)}_j\sim\text{Exp}(\frac{2}{\lambda^2})$. This says that we will achieve $b^{(i)}_{t,j}$ being marginally Laplace distributed by sampling these $s^{(i)}_j$ from the $\text{Exp}(\frac{2}{\lambda^2})$ prior, and then sampling the $b^{(i)}_{t,j}$ from zero-mean Normal distributions with variances $s^{(i)}_j$. Hence
\begin{equation*}
P(b^{(i)}_{t,j}|s^{(i)}_j)=\frac{1}{\sqrt{2\pi s^{(i)}_j}}e^{-[b^{(i)}_{t,j}]^2/[2s^{(i)}_j]},
\end{equation*}
and so $\mathbf{b}^{(i)}_{:,j}$ has the same Normal distribution as $\Phi^{-1}\left\{F_\mathcal{L}[b^{(i)}_{t,j}]\right\}$ but with the variances and covariances scaled up by $s^{(i)}_j$, with $s^{(i)}_j\sim\text{Exp}(\frac{2}{\lambda^2})$. Therefore, also referring back to equation (\ref{fullCovSpec}), it follows that
\begin{align*}
P(\mathbf{b}^{(i)}_{:,j},s^{(i)}_j|\rho^{(i)}_j,\lambda)=&\frac{\lambda^2}{2}e^{-\lambda^2s^{(i)}_j/2}\frac{1}{(2\pi)^{T/2}[s^{(i)}_j]^{1/2}|\boldsymbol{\Sigma}^{(i)}_j|^{1/2}}e^{-{\mathbf{b}^{(i)}_{:,j}}^\top [s^{(i)}_j]^{-1}[\boldsymbol{\Sigma}^{(i)}_j]^{-1}\mathbf{b}^{(i)}_{:,j}/2}.
\end{align*}
To make sampling easier, at this stage we let $s^{(i)}_j=[\nu^{(i)}_j]^{-1}$, leading to the density
\begin{align*}
P(\mathbf{b}^{(i)}_{:,j},\nu^{(i)}_j|\rho^{(i)}_j,\lambda)=&\frac{1}{[\nu^{(i)}_j]^2}\frac{\lambda^2}{2}e^{-\lambda^2/(2\nu^{(i)}_j)}\frac{[\nu^{(i)}_j]^{1/2}}{(2\pi)^{T/2}|\boldsymbol{\Sigma}^{(i)}_j|^{1/2}}e^{-{\mathbf{b}^{(i)}_{:,j}}^\top \nu^{(i)}_j[\boldsymbol{\Sigma}^{(i)}_j]^{-1}\mathbf{b}^{(i)}_{:,j}/2}\\
=&\frac{\lambda^2}{2}e^{-\lambda^2/(2\nu^{(i)}_j)}\frac{[\nu^{(i)}_j]^{-3/2}}{(2\pi)^{T/2}|\boldsymbol{\Sigma}^{(i)}_j|^{1/2}}e^{-{\mathbf{b}^{(i)}_{:,j}}^\top[\boldsymbol{\Sigma}^{(i)}_j]^{-1}\mathbf{b}^{(i)}_{:,j}\nu^{(i)}_j/2},\numberthis\label{bNuPriorFinal}
\end{align*}
where the extra factor of $1/[\nu^{(i)}_j]^2$ is the factor $|d\{[\nu^{(i)}_j]^{-1}\}/d\nu^{(i)}_j|$ due to the change of variable. Assuming that the model will be fit to data standardised to have unit variance, we set the prior on the intercept as $a\sim\mathcal{N}(0,1)$, and we set the prior on the model precision as $\tau_i\sim\text{Gamma}(1,1)$ (which has prior mean 1, with 95\% of the prior mass between 0.025 and 3.7, which we believe is reasonable for these data). Now combining equation (\ref{bNuPriorFinal}) with these prior specifications, and $P(\rho^{(i)}_j|k)=\frac{k}{e^k-1}e^{k\rho^{(i)}_j}$ (for $0\leq \rho^{(i)}_j\leq1$), as well as with the model likelihood (equation (\ref{dataCondParams})), we get:
\begin{align*}
P&(\mathbf{x}_{:,i},\mathbf{B}^{(i)},\boldsymbol{\rho}^{(i)},\boldsymbol{\nu}^{(i)},a_i,\tau_i|\mathbf{X}_{:,\fgebackslash i},\lambda,k)=\left\{\prod_{t=1}^T\prod_{k=1}^{n_t}\sqrt{\frac{\tau_i}{2\pi}}e^{-\tau_i\left(x_{t,i,k}-\mathbf{b}^{(i)}_{t,:}\cdot\mathbf{x}_{t,\fgebackslash i,k}^\top-a_i\right)^2/2}\right\}\\
&\frac{1}{\sqrt{2\pi}}e^{-\left\{\tau_i+a_i^2/2\right\}}\prod_{j=1}^{p-1}\left\{\frac{k}{e^k-1}e^{k\rho^{(i)}_j}\frac{\lambda^2}{2}e^{-\lambda^2/(2\nu^{(i)}_j)}\frac{[\nu^{(i)}_j]^{-3/2}}{(2\pi)^{T/2}|\boldsymbol{\Sigma}^{(i)}_j|^{1/2}}e^{-{\mathbf{b}^{(i)}_{:,j}}^\top[\boldsymbol{\Sigma}^{(i)}_j]^{-1}\mathbf{b}^{(i)}_{:,j}\nu^{(i)}_j/2}\right\}.\numberthis\label{fullJdist}
\end{align*}
Following equation (\ref{fullJdist}), posterior sampling for the model described in Sections \ref{mainModel} and \ref{priorDef} can be implemented through a Gibbs sampler with the steps given in Algorithm \ref{gibbsSampEqs}. We note that Algorithm \ref{gibbsSampEqs} has a relatively low computational cost, because each of the steps (with the exception of step 4) involves sampling from a known distribution for which the parameters can be easily calculated. Then for step 4, we can simply use a slice-sampler to sample $\rho^{(i)}_j$, which has finite support $\rho^{(i)}_j\in[0,1)$. The full derivations of each step of Algorithm \ref{gibbsSampEqs} appear in Supplement B.
\begin{Algorithm}\label{gibbsSampEqs}
A Gibbs sampler with the following steps:
{\small
\begin{enumerate}[label=\textbf{\arabic*})]
\item {\bf Sample:} $a_i$ from: \hfill \makebox[0pt][r]{
\begin{minipage}[b]{\textwidth}
\begin{align*}
P(a_i|\mathbf{x}_{:,i},\mathbf{X}_{:,\fgebackslash i},...)\propto f_\mathcal{N}(a_i|\mu_{a},\sigma_{a})=g_a(a_i),
\end{align*}
\end{minipage}}
\vspace{2ex}\\
where $f_\mathcal{N}$ is the Normal density, $\sigma_{a}^{-2}=1+n\tau_i$ and $\mu_{a}=\sigma_{a}^2\tau_i\sum_{t=1}^T\sum_{k=1}^{n_t}\{x_{t,i,k}-\mathbf{b}^{(i)}_{t,:}\cdot\mathbf{x}_{t,\fgebackslash i,k}^\top\}$.
\vspace*{1ex}
\item {\bf Sample:} $\tau_i$ from: \hfill \makebox[0pt][r]{
\begin{minipage}[b]{\textwidth}
\begin{align*}
P(\tau_i|\mathbf{x}_{:,i},\mathbf{X}_{:,\fgebackslash i},...)\propto\enspace f_\gamma\left(\tau_i|k_\tau,\theta_\tau\right)=g_ \tau(\tau_i),
\end{align*}
\end{minipage}}
\vspace{1ex}\\
where $f_\gamma$ is the density of the gamma distribution with $k_\tau=1+\frac{\sum\limits_{t=1}^Tn_t}{2}$ and
\begin{equation*}
\theta_\tau=1/\{1+\sum\limits_{t=1}^T\sum\limits_{k=1}^{n_t}(x_{t,i,k}-\mathbf{b}^{(i)}_{t,:}\cdot\mathbf{x}_{t,\fgebackslash i,k}^\top-a_i)^2/2\}.
\end{equation*}
\vspace*{-1ex}
\item {\bf Sample:} $\nu^{(i)}_j$ from: \hfill \makebox[0pt][r]{
\begin{minipage}[b]{\textwidth}
\begin{align*}
\quad\quad\quad P(\nu^{(i)}_j|\mathbf{x}_{:,i},\mathbf{X}_{:,\fgebackslash i},...)\propto f_{IG}(\nu^{(i)}_j|\mu_\nu,\lambda_\nu)=g_{\nu_j}(\nu^{(i)}_j),
\end{align*}
\end{minipage}}
\\
where $f_{IG}$ is the density of the inverse Gaussian distribution with parameters\\ $\lambda_\nu=\lambda^2$ and $\mu_\nu=\lambda\bigg/\sqrt{{\mathbf{b}^{(i)}_{:,j}}^\top[\boldsymbol{\Sigma}^{(i)}_j]^{-1}\mathbf{b}^{(i)}_{:,j}}$.
\vspace*{2ex}
\item {\bf Sample:} $\rho^{(i)}_j$ from:\\\vspace*{1ex}\hfill\makebox[0pt][r]{
\begin{minipage}[b]{\textwidth}
\begin{align*}
P(\rho^{(i)}_j|\mathbf{x}_{:,i},\mathbf{X}_{:,\fgebackslash i},...)\propto&\enspace e^{k\rho^{(i)}_j}\frac{1}{|\boldsymbol{\Sigma}^{(i)}_j|^{1/2}}e^{-{\mathbf{b}^{(i)}_{:,j}}^\top\nu^{(i)}_j[\boldsymbol{\Sigma}^{(i)}_j]^{-1}\mathbf{b}^{(i)}_{:,j}/2}=g_{\rho_j}(\rho^{(i)}_j).
\end{align*}
\end{minipage}}
\item {\bf Sample:} $\mathbf{b}^{(i)}_{:,j}$ from: \hfill \makebox[0pt][r]{%
\begin{minipage}[b]{\textwidth}
\begin{align*}
\quad\quad\quad\quad\quad P(\mathbf{b}^{(i)}_{:,j}|\mathbf{x}_{:,i},\mathbf{X}_{:,\fgebackslash i},...)\propto f_\mathcal{N}(\mathbf{b}^{(i)}_{:,j}|\widetilde{\mathbf{m}}^{(i)}_j,\widetilde{\boldsymbol{\Sigma}}_j)=\widetilde{g}_{\mathbf{b}_j}(\mathbf{b}^{(i)}_{:,j}),
\end{align*}
\end{minipage}}
\vspace{2ex}\\
where $f_\mathcal{N}$ is the multivariate Normal density, $[\widetilde{\boldsymbol{\Sigma}}^{(i)}_j]^{-1}=\nu^{(i)}_j[\boldsymbol{\Sigma}^{(i)}_j]^{-1}+[\mathbf{V}^{(i)}_j]^{-1}$,\\and {$\widetilde{\mathbf{m}}^{(i)}_j=\widetilde{\boldsymbol{\Sigma}}^{(i)}_j[\mathbf{V}^{(i)}_j]^{-1}\mathbf{m}^{(i)}_j$}, where the $t^\text{th}$ element of the vector $\mathbf{m}^{(i)}_j$ is
\begin{equation*}
m^{(i)}_{t,j}=\sum_{t=1}^T\sum_{k=1}^{n_t}x_{t,j,k}\left\{x_{t,i,k}-\mathbf{b}^{(i)}_{t,\fgebackslash j}\left(\mathbf{x}_{t,\fgebackslash i,k}\right)_{\fgebackslash j}^\top-a_i\right\}/\sum_{t=1}^T\sum_{k=1}^{n_t}x_{t,j,k}^2,
\end{equation*}
where $\mathbf{b}^{(i)}_{t,\fgebackslash j}$ and $\left(\mathbf{x}_{t,\fgebackslash i,k}\right)_{\fgebackslash j}$ represent $\mathbf{b}^{(i)}_{t,:}$ and $\mathbf{x}_{t,\fgebackslash i,k}$ without the $j\textsuperscript{th}$ elements, respectively, and $\mathbf{V}^{(i)}_j$ is a diagonal matrix, with the $t^\text{th}$ diagonal element equal to $1/\{\tau_i\sum_{t=1}^T\sum_{k=1}^{n_t}x_{t,j,k}^2\}$.
\end{enumerate}
}
\end{Algorithm}

\section{Simulation study}\label{simSec}
\begin{figure}[t]
\vspace{0ex}
\centering\includegraphics[width=0.9\textwidth]{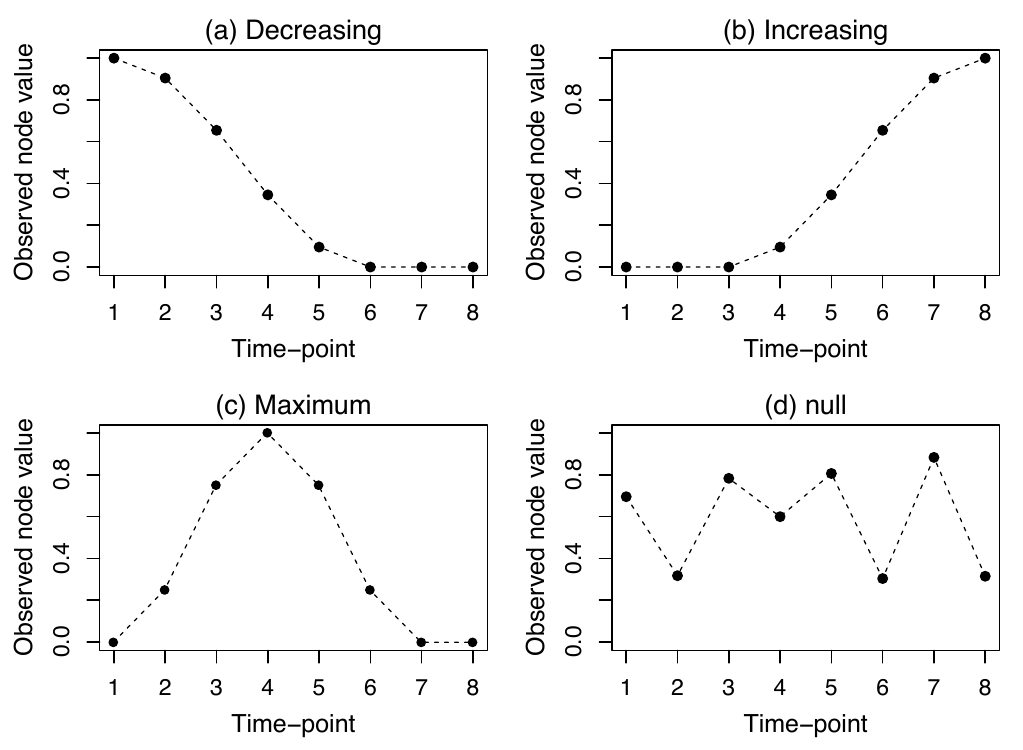}
\vspace{-1ex}
\caption{Simulated time-series of sampled observations at four types of network nodes.}\label{simDataEG}
\vspace{-1ex}
\end{figure}

In this section, we present the results from a simulation study, to test how accurately our model can recover network structure which we know in advance. We generate simulated data with structure that we expect to be typical of real data \citep{nowakowski2017spatiotemporal,mayer2019multimodal}, and then fit the proposed model to the simulated data. To generate the data, the observations $x_{t,i}$ for each node $i$ are generated such that they follow a mean time-series of one of four types (illustrated in Figure \ref{simDataEG}), as follows:
\vspace{-0.5ex}
\begin{enumerate}[label=(\alph*)]
\item Monotonic; decreasing to no signal.
\item Monotonic; increasing from no signal.
\item Maximum: increasing from and decreasing to no signal.
\item Null: random noise.
\end{enumerate}
\noindent
Types (a) and (b) represent node-types of interest to the biological setting, as follows. Type (a) corresponds to genes that are activated (i.e., $x_{t,i}>0$) early in the time-series before becoming de-activated (as we would expect of genes which are important for stem-like cell identity). Type (b) corresponds to genes which only become activated later in the time-series (as we would expect of genes which are important for the identity of mature cells, such as neurons). Types (c) and (d) make the simulated data closer to what we would expect of the real data, by mixing in nodes with other sorts of signals: type (c) corresponds to genes which are active in the middle of the time-series only, and type (d) are null nodes (with random activation). 

\begin{figure}[t!]
\vspace{0ex}
\centering\includegraphics[width=0.7\textwidth]{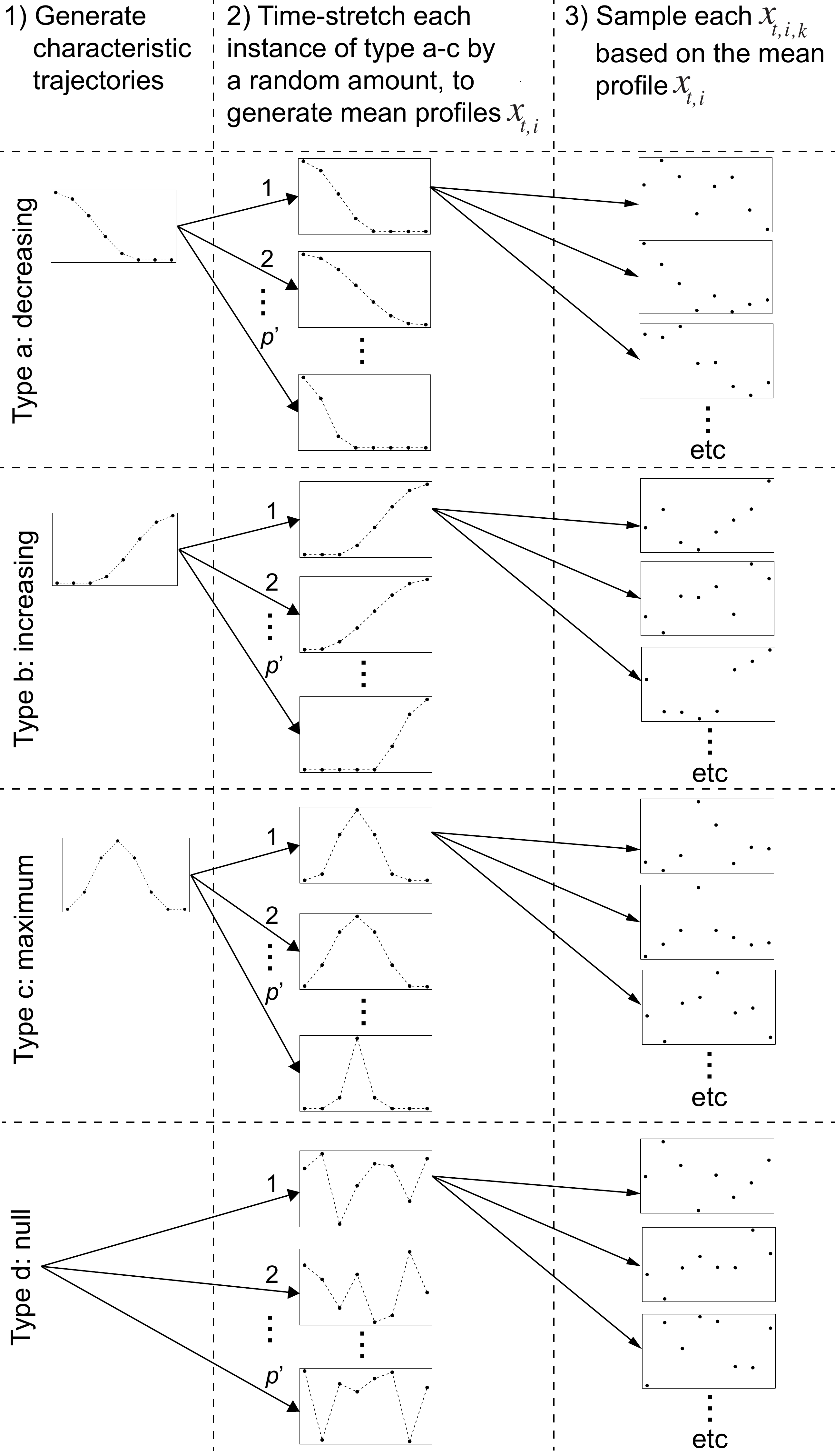}
\vspace{-1ex}
\caption{Overview of the procedure for generating the simulated data.}\label{simDataExplain}
\vspace{-6ex}
\end{figure}

After generating each characteristic mean time-series according to fixed types (a)-(d), we then time-stretch the particular characteristic mean time-series chosen for each node $i$ by a random amount. We do this to reflect the fact that developmental events (as represented by gene-expression measurements) occur at different times in different cells. We achieve this effect by changing the length of the time-series to have a random but uniformly-distributed period $T'\sim\mathcal{U}[T-3,T]$, before zero-padding to return the time-series to its original period $T$. This gives a mean profile $x_{t,i}$, $t=1,...,T$ for each node that is distinct from all other nodes of the same type. We then generate observations $x_{t,i,k}$ for each node according to equation (\ref{mainLinModelEq}) based on these mean profiles, also setting the intercept parameter $a_i$ to 0, and sampling via the Markov chain described in Algorithm \ref{simDataGen}:
\begin{Algorithm}\label{simDataGen}
A Markov chain:\\
{\bf Loop}: $t \text{ in } 1:T$\\
\hspace*{3ex}$X_{t, :, 1} \leftarrow 0 $\hspace*{1ex}// Initialize Markov chain at 0\\
\hspace*{3ex}{\bf Loop}: $r \text{ in } 2:R$\\
\hspace*{6ex}$\mathbf{S} \leftarrow \mathbf{X}_{t, :, r–1}$\\
\hspace*{6ex}{\bf Loop}: $i \text{ in } 1:p$\\
\hspace*{9ex}{\bf Sample}: $S_i \sim N(\mathbf{S}_{\backslash i} . \mathbf{b}^{(i)}_{t,:} , \sigma^2)$\\
\hspace*{6ex}{\bf end loop}\\
\hspace*{6ex}$\mathbf{X}_{t, :, r} \leftarrow \mathbf{s} $\\
\hspace*{3ex}{\bf end loop}\\
{\bf end loop}
\end{Algorithm}
\noindent
where $\mathbf{X}$ is a $T \times p \times R$ array containing the sampled data, $\mathbf{S}$ is a vector of length $p$ which temporarily stores intermediate results, and the elements of $\mathbf{b}^{(i)}$ are specified as:
\begin{equation*}
b^{(i)}_{t,j}=
\begin{cases}
\vspace*{1ex}
1/{p'}, & \text{if nodes $i$ and $j$ are of the same type}\  \\
\vspace*{1ex}
0, & \text{otherwise}\\
\end{cases}
\end{equation*}
where $p'$ is the number of nodes $j$ of the same type as $i$. The number of MCMC samples in the Markov chain specified in Algorithm \ref{simDataGen} is given by the variable $R$: we use $R=10^4$, and after thinning to take one sample in every 100, we choose the final 25 (thinned) samples to pass forward to the model fitting after adding the mean characteristic profiles. That is, we have 25 samples per time-point, i.e., $n_t=25$, where autocorrelation analysis and an experimentation with burn-in times show no evidence against them being independently and identically distributed at each time-point group. We note that in the simulation we specify means that vary with time but our model has constant mean. In practice this does not make any difference as the data are always standardised before model fitting, but the user can easily make the intercept time-varying if this is a concern. The procedure for generating the simulated data is also illustrated in Figure \ref{simDataExplain}.

We generate each time-series with $T=8$, $n_t=25$ (constant for all values of $t$), and $p'=10$, adding noise with standard deviations $\tau_i^{-1/2}=\tau^{-1/2}\in\{0.1,0.2,0.3\}$. Then, we apply Algorithm \ref{gibbsSampEqs}, and calculate each $\hat{b}^{(i)}_{t,j}$ from the median of the corresponding posterior. We infer an edge between nodes $i$ and $j$ if $\hat{b}^{(i)}_{t,j}\neq0$, after thresholding the $\hat{b}^{(i)}_{t,j}$ to remove trivially small values, i.e., if $|\hat{b}^{(i)}_{t,j}|\geq\phi$. We generate ROC (receiver-operator characteristic) curves as this threshold $\phi$ is decreased to 0 from $\text{max}|\hat{b}^{(i)}_{t,j}|$ (for $t\in\{1,...,T\}$ and all $j$). We generate these curves from the true-positives (TP) and false-positives (FP) which we calculate from the ground-truth network edges $b^{(i)}_{t,j}$ and estimated network edges $\hat{b}^{(i)}_{t,j}$ as follows:
\vspace{-2ex}
\begin{align*}
|\hat{b}^{(i)}_{t,j}|>0\enspace\text{for}\enspace|b^{(i)}_{t,j}|>0\implies\enspace&\text{TP}\\
\text{and}\enspace\enspace|\hat{b}^{(i)}_{t,j}|>0\enspace\text{for}\enspace|b^{(i)}_{t,j}|=0\implies\enspace&\text{FP}
\end{align*}
We generate an average ROC curve over 1000 repetitions of this procedure, and then calculate an AUC (area under curve) statistic for this average ROC curve. 

\begin{figure}[t]
\vspace{0ex}
\centering\includegraphics[width=0.9\textwidth]{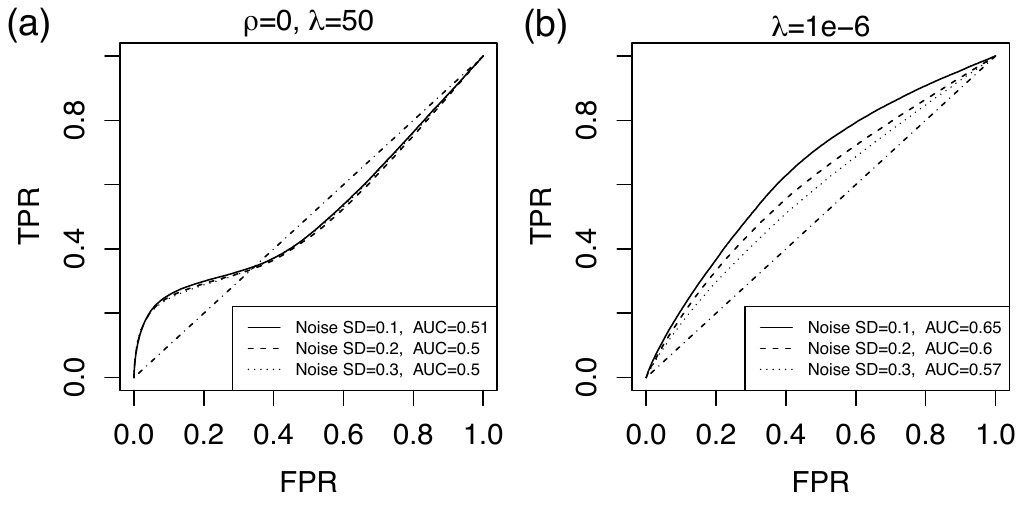}
\vspace{-1ex}
\caption{Accuracy of network inference by the model without either sparsity within, or across, time. (a) Model performance when sparsity across time is removed. (b) Model performance when sparsity within time is removed. Abbreviations: TP, true positives; FP, false positives.}\label{simDataNullParam}
\vspace{-1ex}
\end{figure}

\begin{figure}[t!]
\vspace{0ex}
\centering\includegraphics[width=0.9\textwidth]{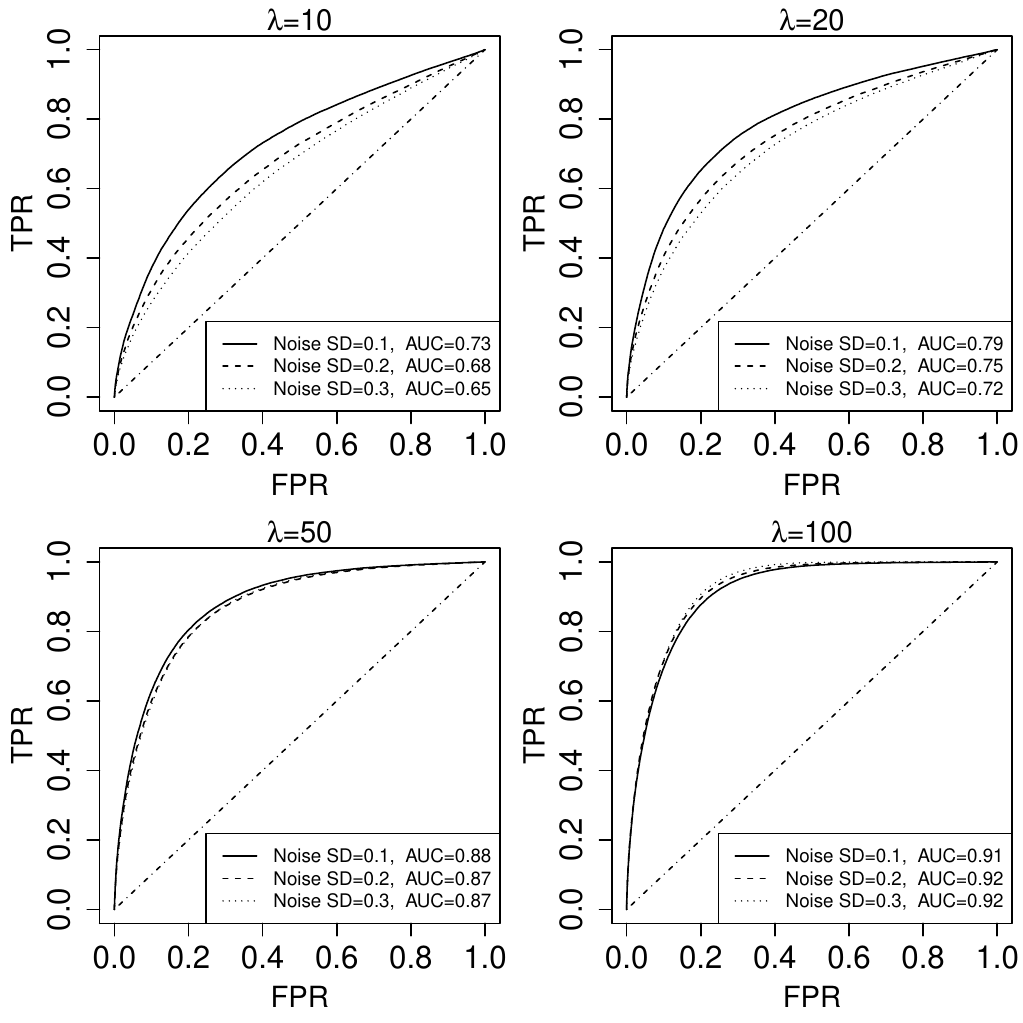}
\vspace{-1ex}
\caption{Accuracy of network inference by the model with both sparsity within and across time. Abbreviations: TP, true positives; FP, false positives.}\label{simDataRes}
\vspace{-1ex}
\end{figure}
We assess the performance of our full model using both sparsity within and sparsity across time, compared with the scenarios when one of these priors is excluded from the model. To exclude sparsity across time, we enforce $\rho=0$, and to exclude sparsity within time, we use $\lambda\rightarrow0$: these results are shown in Figure \ref{simDataNullParam}. With $\rho=0$, there is no correlation of the model parameters across time, and so we see the effect of inferring the networks separately for each time-point; i.e., sparsity across time is removed: in this case, $\text{AUC}=0.5$ indicates that none of the intended structure in the data is being detected. Alternatively, as $\lambda\rightarrow0$, the prior becomes flat or uninformative, and so in this case we see the effect of fitting the model without the sparsity within time. We again note that decoupling these types of sparsity is made possible by design with the model structure we propose, unlike alternatives such as Laplace-NEG \citep{shimamura2016bayesian}. Then for the full model (which includes the priors to enforce both the sparsity within time and across time), we repeated the simulation for various values of sparsity parameter $\lambda$: Figure \ref{simDataRes} shows the results (with hyperparameter $k=20$, equivalent results with $k=10$ and $k=50$ are shown in Figures S10 and S11 in Supplement D). When we include the priors for both sparsity within and across time, we can achieve AUC of 0.9 or more, as long as the sparsity parameter $\lambda$ is large enough. This result demonstrates that our priors are responsible for good detection of network edges with respect to the ground-truth in these simulated data. We also found that these results were not very sensitive to $p'$, the number of covariates included in the simulated data. Figure S12 shows equivalent results to Figure \ref{simDataRes}, except with the number of covariates halved to $p'=5$. In this case we found that the network inference is a bit more accurate, as would be expected with a smaller number of variables to predict; although the difference is minimal as long as the sparsity is great enough.

\begin{figure}[t!]
\vspace{0ex}
\centering\includegraphics[width=0.9\textwidth]{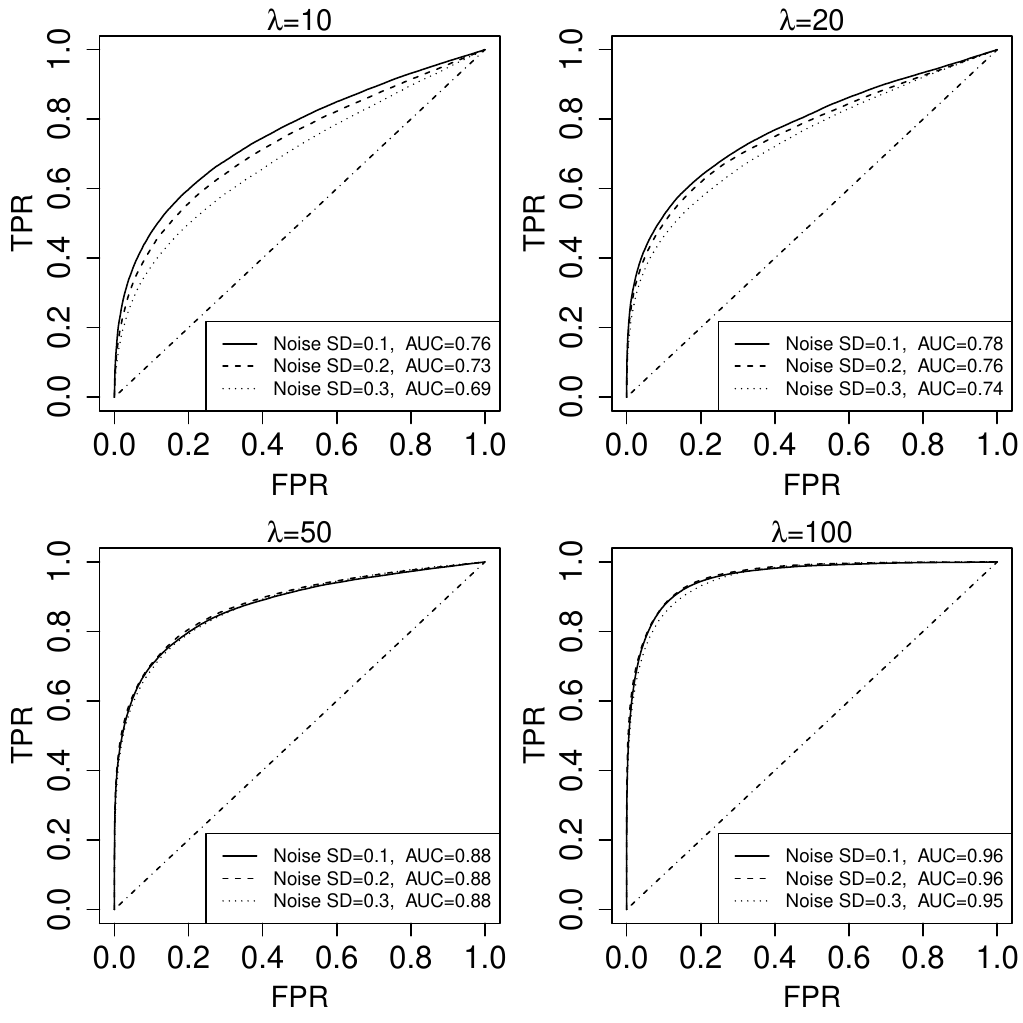}
\vspace{-1ex}
\caption{Accuracy of network inference in the simulation study with 60\% dropouts (with $k=20$). Abbreviations: TP, true positives; FP, false positives.}\label{simDataResDrop}\vspace{-1ex}
\vspace{-1ex}
\end{figure}

Dropouts, or missing values, are a well known source of technical noise in single-cell transcriptome data. These missing values are replaced by zeros, leading to `zero inflation'. A characteristic of this dropout effect is that data-values which are already small are more likely to drop out (i.e., get missed out), than values which are larger in magnitude. This is data missing not-at-random, an effect that can be challenging to model \citep{kharchenko2014bayesian}. Dropout rates (i.e., the proportion of data-values missing from the data-set) are often over $60\%$ in typical single-cell transcriptome data-sets that we have seen, such as the one analysed in Section \ref{resSec}. To test the robustness of our method to dropouts, we used a well known and effective model of the dropout effect, published previously by \cite{pierson2015zifa}. This model of dropouts specifies the probability of an observed data-value dropping out as
\begin{equation}
p_{t,i,k}=\exp(-\omega\tilde{x}^2_{t,i,k}),\label{dropEq}
\end{equation}
where the parameter $\omega$ controls the dropout rate (decreasing $\omega$ increases the number of dropouts), and $\tilde{x}_{t,i,k}$ is the data-value that would have been present without the dropout effect. This is essentially a hurdle model, with $h_{t,i,k}\sim\text{Bernouilli}(p_{t,i,k})$, so that the observed data $x_{t,i,k}$ (i.e., with dropouts included) is modelled as $x_{t,i,k}=h_{t,i,k}\cdot\tilde{x}_{t,i,k}$. We found that under this model, $\omega=2$ leads to a dropout rate of around $66\%$ in data-sets generated according to the data-simulation procedure presented earlier in this section. We used this value of $\omega=2$, and repeated our simulation study now with the addition of this dropout effect, carrying out out the same ROC-curve analysis as before. The results of this analysis (again with $\lambda=50$ and $k=20$) are shown in Figure \ref{simDataResDrop}: we found that our method is quite resilient to dropouts, with only a moderate decrease in performance compared to the results shown in Figure \ref{simDataRes}. The time-varying aspect of the model apparently helps to maintain performance when many dropouts are present, because when some values in the time-series are missing, sparsity across time encourages interpolation over the missing values. Interestingly, in very sparse cases, the dropout effect may even be helpful, possibly via a de-noising mechanism, as follows. Referring to the generative model of equation (\ref{dropEq}), it's clear that the dropouts mostly take place for small values of $x_{t,i,k}$. As these are much more likely to correspond to noise than larger values do, this leads to a strong de-noising effect. Finally, we note that to include a hurdle model or dropout effect in a model likelihood such as the one proposed in Section \ref{mainModel} would result in a much more computationally intensive model fitting procedure than the one we propose in our Algorithm \ref{gibbsSampEqs} of Section \ref{postInf}.

\section{Single-cell gene-expression data}\label{resSec}
In this section we present an example application of our proposed methodology, to single-cell gene-expression data. These data have been published previously by \cite{nowakowski2017spatiotemporal}, and are publicly available from the NCBI database of genotypes and phenotypes (dbGaP), under accession number phs000989.v3\\
In this context, $x_{t,i,k}$ represents the log-expression of gene $i$, defined as log(transcript counts + 1), in sample $k$ from time $t$. For the pseudo-time assignments for each cell, we use cell-type classifications provided with the data, together with an ordering for these cell-types according to the developmental lineage (for full details see Supplement C). We fitted the model to $n=1557$ cell samples, and $p=22988$ genes/nodes, reduced to $p=212$ for each individual model fit by variable screening. For the fitting we used values of $\lambda=20$ and $k=1$: these values were chosen by grid-search stochastic EM (Figure S13 in Supplement D). Fitting the model as described, we obtained posterior distributions for each model parameter $b^{(i)}_{t,j}$, and we used the posterior medians as posterior summaries, $\hat{b}^{(i)}_{t,j}$. To fit each model, we ran the Gibbs' sampler proposed in Algorithm \ref{gibbsSampEqs} for $1\times10^4$ samples (after $1\times10^3$ samples burn-in), which took 1.4 hours for each target-node on one core of a Macbook Pro laptop (mid 2015, 2.8 GHz, 16GB RAM).

\begin{figure}[t]
\vspace{0ex}
\centering\includegraphics[width=0.99\textwidth]{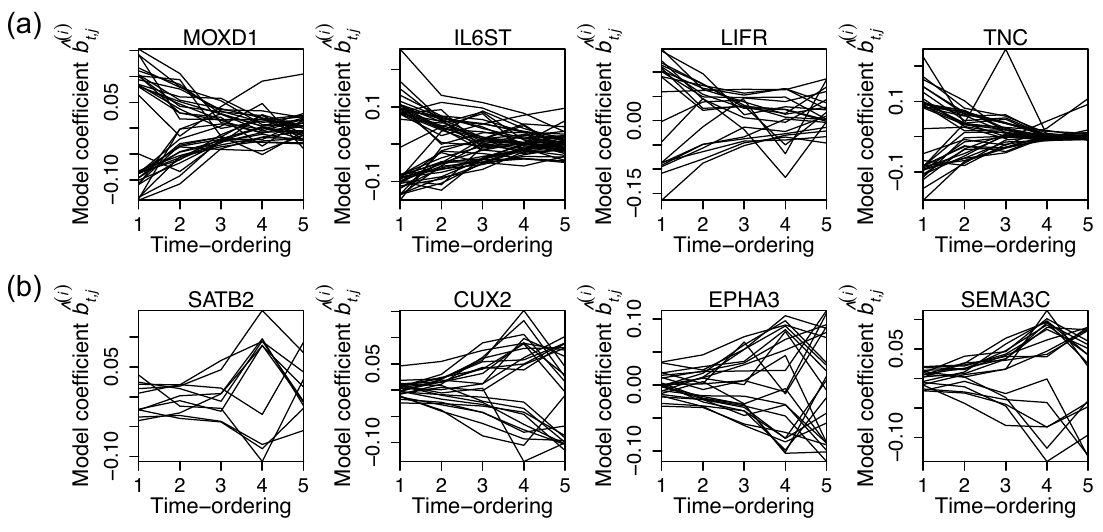}
\vspace{-1ex}
\caption{Inferred model parameters $\hat{b}^{(i)}_{t,j}$, for genes characteristic of: (a) stem-cells; (b) mature cells (neurons). Non-zero parameters $\hat{b}^{(i)}_{t,j}$ infer the local network structure around gene/node $i$. Parameters which are zero for every time-point are not plotted.}\label{bestRegs}
\vspace{-1ex}
\end{figure}

The model was fitted initially to a panel of 25 genes, as target-nodes: these genes were chosen in an unbiased way by searching the biological sciences literature for genes that are important in this biological setting, and then analysing those that were present in this data-set after quality control. Estimated model parameters $\hat{b}^{(i)}_{t,j}$ for a selection of these genes are shown in Figure \ref{bestRegs}, and the full panel is shown in Figures S14 and S15.

\begin{figure}[t]
\vspace{-0ex}
\centering\includegraphics[width=0.9\textwidth]{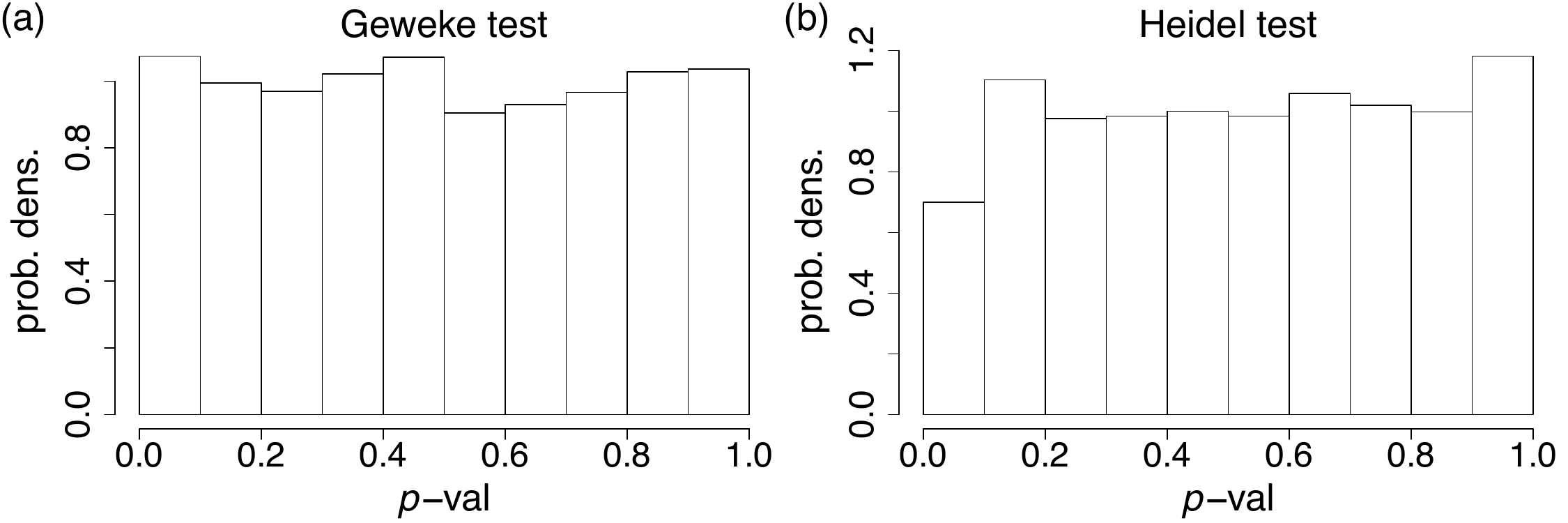}
\vspace{-1ex}
\caption{Convergence-test results, for the parameters $b^{(i)}_{t,j}$ which appear in Figures \ref{bestRegs} and S14 and S15.}\label{convPlotPs}
\vspace{-1ex}
\end{figure}

\begin{figure}[t]
\vspace{-0ex}
\centering\includegraphics[width=0.5\textwidth]{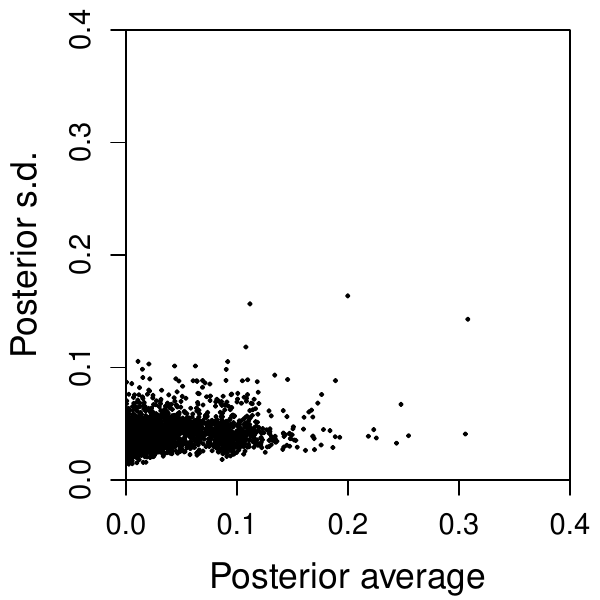}
\vspace{-1ex}
\caption{Estimates of the spread of the posterior distributions for the parameters $b^{(i)}_{t,j}$ which appear in Figures \ref{bestRegs} and S14 and S15.}\label{postSDs}
\vspace{-1ex}
\end{figure}

We carried out Geweke \citep{geweke1991evaluating} and Heidel \citep{heidelberger1981spectral} convergence tests, using the {\tt R} package {\tt CODA} \citep{plum2006coda}, for the sampler outputs for all the parameters $b^{(i)}_{t,j}$ shown in Figures \ref{bestRegs}, S14 and S15. These convergence test results appear in Figure \ref{convPlotPs}. In convergence tests such as these, if an individual $p$-value is significantly small, it can be taken as evidence that the chain has not yet converged. Hence, the uniform distributions of $p$-values shown in Figure \ref{convPlotPs}, in which these $p$-values are aggregated over all the test results, indicate that the MCMC sampler has converged for these target-nodes. Then, to give an indication of how `stiff' or `sloppy' these parameters are, we estimated the standard-deviations of these posterior distributions for this panel of genes: these are plotted against the corresponding posterior averages in Figure \ref{postSDs}. These posterior standard deviations are typically much smaller in magnitude than the posterior averages, demonstrating that the posteriors are not `sloppy', and indicating that the estimates from our model are reliable. We also wanted to make sure that our results are not driven by a few outlier cells. So we repeated the inference for this same panel of genes, but now using only a random sample of 50\% of the cells originally used, i.e., $n=779$. The results of this analysis are plotted in Figure S16, for the same genes as are shown in Figure \ref{bestRegs}. The results shown in these figures are clearly very similar, and therefore we conclude that our results here are not driven by outliers.

Figure \ref{bestRegs}a shows inferred model parameters $\hat{b}^{(i)}_{t,j}$, for a selection of nodes/genes which are characteristic of stem cells, and of neurons (i.e., mature cells), selected from the full panel of 25 genes. We expect stem cells to predominate at earlier times, and hence we expect to see decreasing time-series for genes which are characteristic of this type of cell. On the other hand, we expect mature cells such as neurons to predominate at later times, and so we expect to see increasing time-series for genes characteristic of this type of cell. As would be expected for stem-cell genes, important model parameters $b^{(i)}_{t,j}$ tend to decrease in magnitude during the developmental trajectory as cells go from stem-cell to mature cell types (e.g., gene transcript MOXD1). Figure \ref{bestRegs}b then shows, as would be expected, that important model parameters $\hat{b}^{(i)}_{t,j}$ become non-zero (corresponding to network edges appearing) late in the developmental trajectory, when the cells become neurons and hence their characteristic gene regulatory program is activated (e.g., for SATB2). Equivalent results to Figure \ref{bestRegs}a-b for the full panel of 25 genes analysed then appear in Figures S14 and S15 respectively in Supplement D. In these figures, we also see similar results: for genes that tend to be active in stem-cells, model parameters $b^{(i)}_{t,j}$ tend to decrease in magnitude during the developmental trajectory as cells go from stem-cell to mature cell types (Figure S14), and {\it vice-versa} for genes which are important to mature cells such as neurons (Figure S15). 

We wish to infer a network edge between nodes $i$ and $j$ if $|\hat{b}^{(i)}_{t,j}|>0$. We estimate these $\hat{b}^{(i)}_{t,j}$ from the posterior medians, but because we find that many of these medians are close to, but not exactly zero, we set $\hat{b}^{(i)}_{t,j}$ to zero in such cases by thresholding. Therefore, we infer `no edge' between nodes $i$ and $j$ when the posterior median is close to zero. Hence, if (and only if) $|\hat{b}^{(i)}_{t,j}|>\phi$, where $\phi$ is the threshold parameter, we would infer a network edge between nodes $i$ and $j$ at time $t$ (for the model fit around node $i$). We note that the local model fitting (equation (\ref{mainLinModelEq})) does not depend on this network estimation. Hence, this thresholding can take place independently of the computationally-intensive MCMC sampling. Thus, we leave $\phi$ as a tuning parameter, which can be varied by the user in real time to interpret results, equivalently to changing the resolution or granularity in a visualisation. We recommend the user does a full sweep through $\phi\in[0,\infty]$ to interpret the results. We also note that if $|\hat{b}^{(j)}_{t,i}|>\phi$ (for the independent model fit around node $j$ rather than node $i$), we would independently infer an edge between nodes $i$ and $j$ at time $t$. Thus, some inconsistency may arise, due to these independent model fits around nodes $i$ and $j$. To deal with this, we use the `min\_symmetrisation' scheme of \cite{kolar2010estimating}, inferring an edge between nodes $i$ and $j$ at time $t$, i.e., $\hat{A}_{i,j,t}\neq0$, if and only if $|\hat{b}^{(i)}_{t,j}|>\phi$ and $|\hat{b}^{(j)}_{t,i}|>\phi$.

\begin{figure}[t]
\vspace{0ex}
\centering\includegraphics[width=0.8\textwidth]{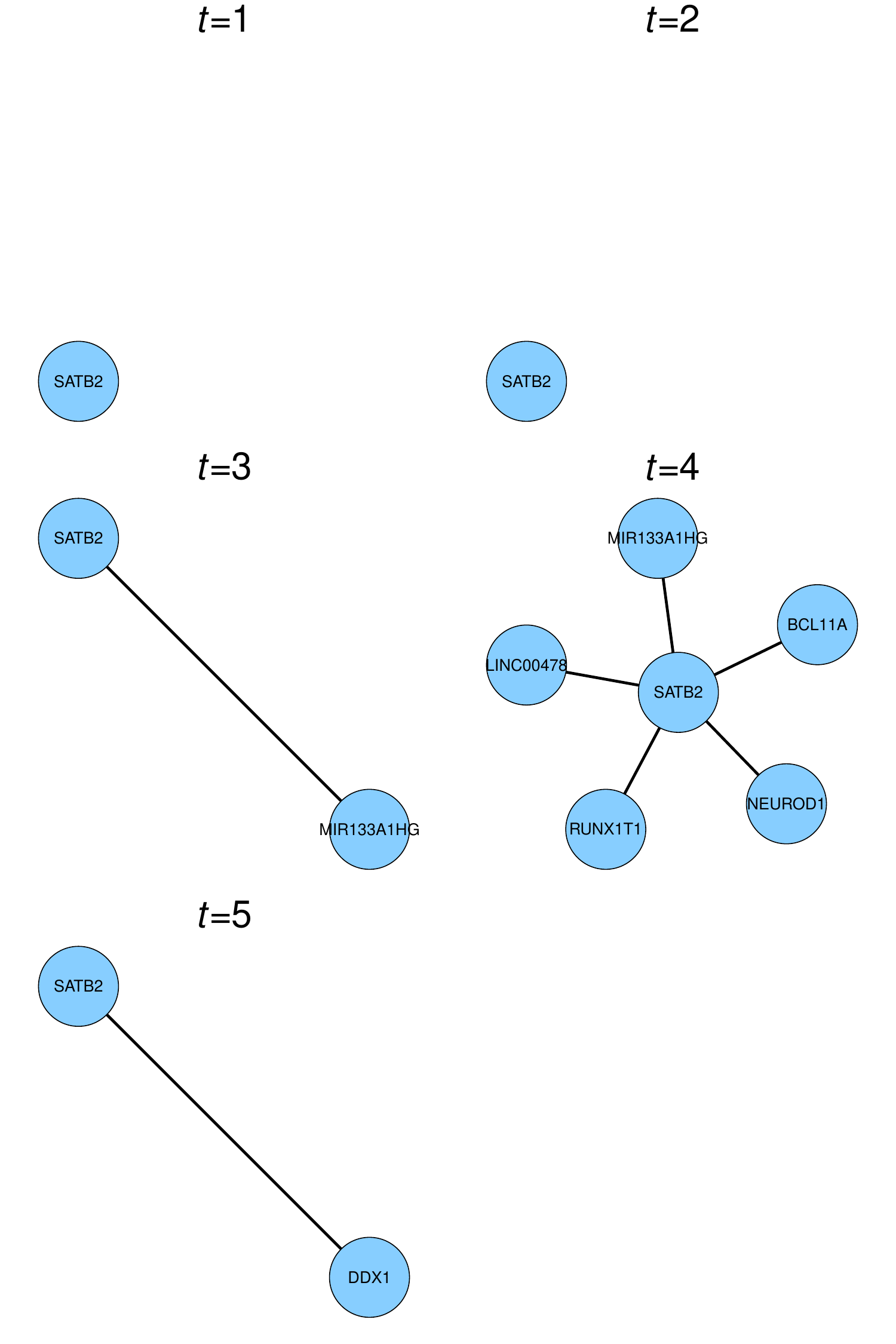}
\vspace{-1ex}
\caption{Time-varying network structure inferred around the gene SATB2. This gene is characteristic of certain types of neuron, and hence we would expect network structure to appear at later times, when the cell type-specific gene regulatory program becomes activated.}\label{SATB2netPlot}
\vspace{-6ex}
\end{figure}

Plots of the inferred network structure around an example of a gene shown in Figure \ref{bestRegs}b, namely SATB2, are shown in Figure \ref{SATB2netPlot}, after `min\_symmetrisation' (Section \ref{modelOverviewSec}) with $\phi=0.05$. In addition to the neuronal identity gene SATB2 \citep{alcamo2008satb2}, several of the genes shown in Figure \ref{SATB2netPlot} are already known to be important in neuronal development, including NEUROD1 which initiates the programme of neuronal development \citep{pataskar2016neurod1} and RUNX1T1 which regulates the differentiation of neurons from neural stem cells \citep{linqing2015runx1t1}, as well as the neuronal circuit-formation gene BCL11A \citep{john2012bcl11a}. Intriguingly, this network structure also includes MIR133A1HG and LINC00478, which are (respectively) examples of micro-RNA (miRNA) and long non-coding RNA (lncRNA). Non-coding RNA transcripts such as these do not get translated into proteins, as would usually be the case for a transcript from a region of DNA which codes for a gene. Instead, non-coding RNA transcripts are known to play an important role in gene regulation \citep{cech2014noncoding}. However, we still only understand a small amount about their function, and gene regulation involving these sorts of non-coding RNA is an important research topic. We note that MIR133A1HG and LINC00478 are promising candidates for for further experimental investigation which have been identified using our proposed methodology.

\clearpage

Next, we compared our method with alternative network inference method for single-cell transcriptome data. Methods for inferring time-varying network structure in data of this type include alternatives designed for many fewer nodes than our method can handle, such as the work of \cite{matsumoto2017scode}, which is designed to infer structure in networks with fewer than 100 nodes. However, there is also a static network inference method available for single-cell transcriptome data called `SCENIC' \citep{aibar2017scenic}, that can be used to infer structure in large networks of 20000 or more nodes, and that is therefore also appropriate for the data-set analysed here. For comparison with our proposed methodology, we ran the SCENIC method on the same single-cell gene-expression data-set already analysed. Equivalently to our proposed method, SCENIC returns fitted model parameters that indicate the strength of the network connection between a pair of nodes, or genes: maintaining equivalent notation, we label these SCENIC model parameters $b^{\prime(i)}_j$. Thus, by again choosing a threshold $\phi^\prime$, it is possible to infer network structure by inferring edges between the pair of nodes $i$ and $j$ if the corresponding fitted model parameter $|\hat{b}^{\prime(i)}_j|>\phi^\prime$. We choose $\phi^\prime$ so as to maintain the same number of connections to each target in the network inferred by the SCENIC method, as compared with our method. Table \ref{SCENICtab} shows the genes inferred in the network structure around the neuronal identity gene SATB2 (summarised from Figure \ref{SATB2netPlot}), together with the genes equivalently found from the SCENIC method (setting $\phi^\prime$ to maintain the same number of connections). 

\begin{table}
{
\centering
\begin{tabular}{l|l}
Our method & SCENIC   \\
\hline
BCL11A & ARPP21   \\
DDX1 & CHL1     \\
LINC00478 & KIAA1598 \\
MIR133A1HG & MEF2C    \\
NEUROD1 & NFIA     \\
RUNX1T1 & RUNX1T1 
\end{tabular}
\caption{Comparison of nodes inferred in static network structure by our method, and the SCENIC method.}\label{SCENICtab}
}
\end{table}

Of the genes shown in Table \ref{SCENICtab}, just as with those found by our method, those found by the SCENIC method are mostly already known to be involved in neural development, as follows. ARPP21 is involved with branching of dendrites \citep{rehfeld2018rna}, CHL1 and KIAA1598 are thought to be involved in neuronal migration and axon formation \citep{alsanie2017homophilic,toriyama2006shootin1}, and MEF2C and NFIA are known to be important for neural stem and progenitor cell differentiation \citep{li2008transcription,piper2010nfia}. However, we note that the SCENIC method is not able to infer time-varying network structure, as our method can: to make the comparison shown Table \ref{SCENICtab}, the time-varying aspect of the network structure inferred by our method had to be `flattened out'.

It is challenging to visualise in a meaningful way the entire structure of a large network, such as the full genome-wide network inferred here, if it is inferred for all 22989 nodes. This challenge becomes even larger when the dimension of time is added. After fitting the
\begin{figure}[t!]
\vspace{0ex}
\centering\includegraphics[width=0.8\textwidth]{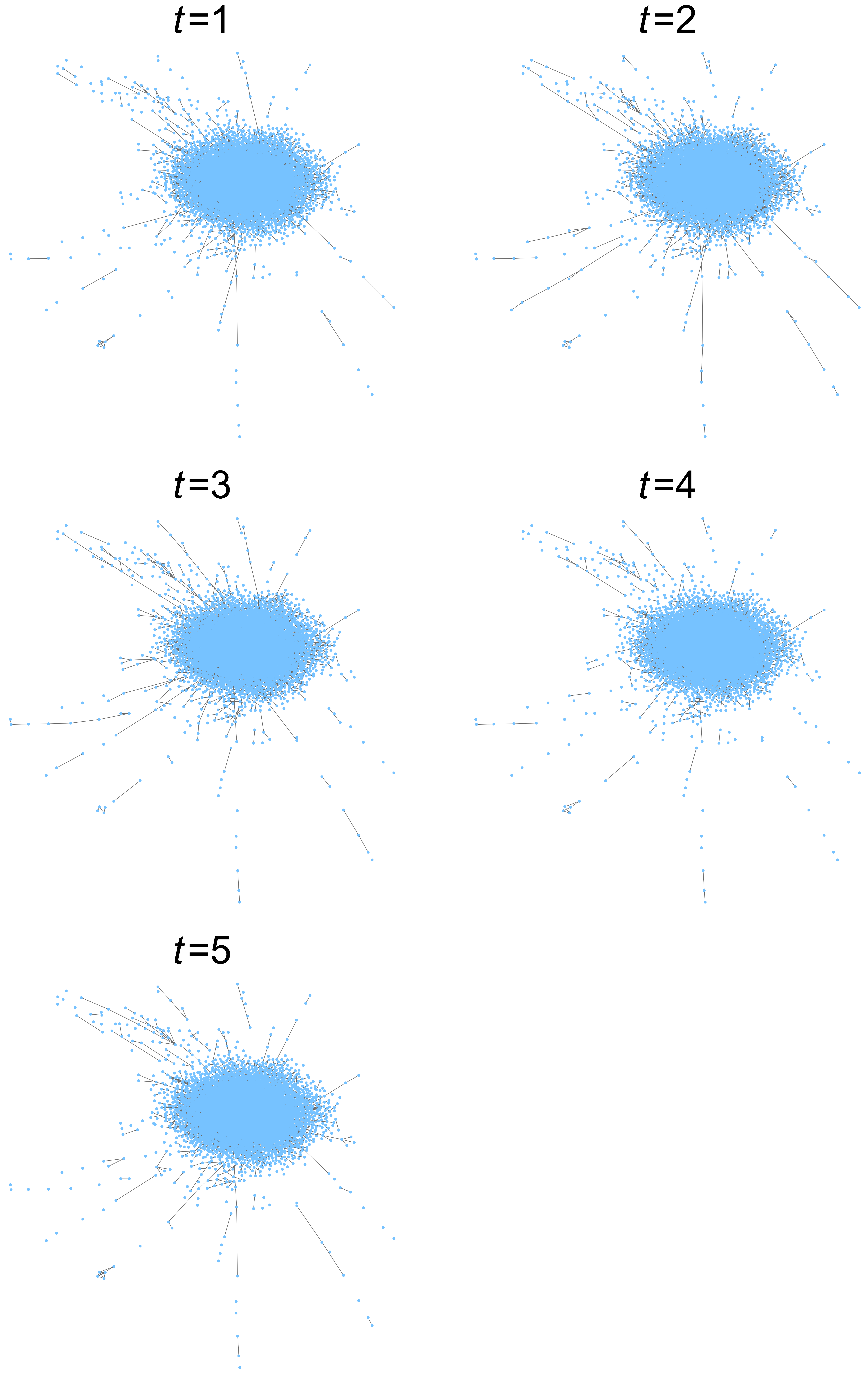}
\vspace{-1ex}
\caption{Time-varying network structure of the fully-connected component (11133 nodes) of the inferred genomic network.}\label{fullNetPlot}
\vspace{-6ex}
\end{figure}
\clearpage
\noindent
model to all 22989 target-nodes on a high-performance computing cluster, we inferred the structure of this network, and found the fully connected component (11133 nodes), which is shown in Figure \ref{fullNetPlot}. As a minimum, it can be seen from this figure that the network structure changes gradually rather than suddenly with time, as we would expect from our proposed methodology.

\section{Discussion}\label{discussSect}
In this paper, we have proposed a new model to infer time-varying network structure. This model makes use of a novel prior structure we introduce here, which extends the Bayesian lasso to the time-varying case. The novel structure of this prior allows for effective modelling of time-varying network structure even in situations where there are very few time-points, as is typical in cell-biological (i.e., `omics) data. We also found that the model fitting and inference procedure we have proposed works well even in with large networks of over 20000 nodes, which compares very well with alternatives (see for example the work by \cite{matsumoto2017scode}).

We used simulated data to assess the ability of the proposed model to accurately infer time-varying network structure, and we showed that the model is effective in inferring time-varying genomic network structure from single-cell gene-expression data. However, we note that genomic network structure which is inferred from only gene-expression data (as we do here) is not guaranteed to correspond to true gene regulatory patterns. To strengthen any belief that the inferred genomic network structure corresponds to true gene regulatory patterns rather than simply gene co-expression patterns, evidence from, for example, chromatin binding and epigenomic data could also be incorporated into the model \citep{novershtern2011physical}. We intend to incorporate such data as the next stage of the development of this model. Specifically, we will do this by allowing the sparsity parameter $\lambda$ to vary for each pair of nodes $i$ and $j$, depending on any prior evidence of a physical interaction between the protein-product of gene $j$ with the DNA or surrounding chromatin of gene $i$.

Another characteristic of the single-cell transcriptome data analysed here is that the data are zero-inflated. This is a case of data missing-not-at-random, because the dropout events which lead to the extra zeros in the data are more likely to occur when the true transcriptome level is low \citep{kharchenko2014bayesian}. As part of the next stage of the development of this model, we intend to account for dropouts as other authors have done \citep{van2017magic}, for example by explicitly including the dropout events in the model likelihood \citep{pierson2015zifa}. We also note that existing time-inference methods for data such as those presented here are algorithmic, rather than model-based. Hence it is not easy to obtain uncertainties on the inferred times when using these methods. Thus, we would like to develop a model-based time-inference method that will provide such uncertainties, and then feed these uncertainties directly into the time-varying network model we have proposed. We also note that in other contexts, it could complicate the analysis if there is uneven time-sampling. For example, if we expect highly deterministic behaviour with little noise, but have data with time-sampling at known but uneven time-points, the method might need to be adapted. Specifically, in that context we would expect to see larger changes in parameters over larger time-intervals: this structure is not explicitly captured by our model, in its current form.

Understanding interactions between genes and their transcriptional regulators is a fundamental question in genomics, and network models are a natural way to represent and analyse groups of interactions between genes and their regulators. Biomedical science in the high-throughput genomic age has been developing ever more innovative ways to collect increasingly vast quantities of data. However, the statistical techniques to represent, analyse and interpret such data still lag behind the means to generate them. In particular, there is currently a lack of good computational statistical methodology to represent and analyse changes in gene-regulatory interactions as cells are specified and change state - an issue we address with the time-varying network model that we propose here. The computational-statistical tools that we are developing allow novel characterisation of genomic interactions in important settings, adding to knowledge of fundamental biological principles, and motivating further investigation by targeted experiments.

\section*{Acknowledgements}
We are grateful to Aaron Diaz, Tom Nowakowski, Alex Pollen, and Aparna Bhaduri, for helpful discussions, insightful comments, and useful advice throughout this project, and for providing early access to the data. The work of the first author was supported by the MRC grant MR/P014070/1. The work of the second and third author was partially supported by The Alan Turing Institute under the EPSRC grant EP/N510129/1.

\begin{supplement}[id=supp]
  \sname{Supplement}
  \stitle{Supplementary Information}
   \sdatatype{Appears below}
  \sdescription{Derivations; Details of data pre-processing; Supplementary Figures S1-S14.}
\end{supplement}
  
\begin{supplement}[id=soft]  
    \sname{Supplement}
  \stitle{Software}
   \sdatatype{Online respository}
  \sdescription{An R package containing an efficient implementation of the model proposed in this paper can be installed in R by typing: {\tt install.packages("devtools")} and then: {\tt devtools::install\_github("tombartlett/SBDN")}\\ 
  This package contains an R function which calls a C++ implementation of the Gibbs sampler described in Algorithm \ref{gibbsSampEqs}.}  
\end{supplement}

\bibliography{references.bib}

\begin{thebibliography}{51}

\bibitem[\protect\citeauthoryear{Aibar et~al.}{2017}]{aibar2017scenic}
\begin{barticle}[author]
\bauthor{\bsnm{Aibar},~\bfnm{Sara}\binits{S.}},
  \bauthor{\bsnm{Gonz{\'a}lez-Blas},~\bfnm{Carmen~Bravo}\binits{C.~B.}},
  \bauthor{\bsnm{Moerman},~\bfnm{Thomas}\binits{T.}},
  \bauthor{\bsnm{Imrichova},~\bfnm{Hana}\binits{H.}},
  \bauthor{\bsnm{Hulselmans},~\bfnm{Gert}\binits{G.}},
  \bauthor{\bsnm{Rambow},~\bfnm{Florian}\binits{F.}},
  \bauthor{\bsnm{Marine},~\bfnm{Jean-Christophe}\binits{J.-C.}},
  \bauthor{\bsnm{Geurts},~\bfnm{Pierre}\binits{P.}},
  \bauthor{\bsnm{Aerts},~\bfnm{Jan}\binits{J.}}, \bauthor{\bparticle{van~den}
  \bsnm{Oord},~\bfnm{Joost}\binits{J.}} \betal{et~al.}
(\byear{2017}).
\btitle{SCENIC: single-cell regulatory network inference and clustering}.
\bjournal{Nature methods}
\bvolume{14}
\bpages{1083}.
\end{barticle}
\endbibitem

\bibitem[\protect\citeauthoryear{Alcamo et~al.}{2008}]{alcamo2008satb2}
\begin{barticle}[author]
\bauthor{\bsnm{Alcamo},~\bfnm{Elizabeth~A}\binits{E.~A.}},
  \bauthor{\bsnm{Chirivella},~\bfnm{Laura}\binits{L.}},
  \bauthor{\bsnm{Dautzenberg},~\bfnm{Marcel}\binits{M.}},
  \bauthor{\bsnm{Dobreva},~\bfnm{Gergana}\binits{G.}},
  \bauthor{\bsnm{Fari{\~n}as},~\bfnm{Isabel}\binits{I.}},
  \bauthor{\bsnm{Grosschedl},~\bfnm{Rudolf}\binits{R.}} \AND
  \bauthor{\bsnm{McConnell},~\bfnm{Susan~K}\binits{S.~K.}}
(\byear{2008}).
\btitle{Satb2 regulates callosal projection neuron identity in the developing
  cerebral cortex}.
\bjournal{Neuron}
\bvolume{57}
\bpages{364--377}.
\end{barticle}
\endbibitem

\bibitem[\protect\citeauthoryear{Alexander
  et~al.}{2009}]{alexander2009understanding}
\begin{barticle}[author]
\bauthor{\bsnm{Alexander},~\bfnm{Roger~P}\binits{R.~P.}},
  \bauthor{\bsnm{Kim},~\bfnm{Philip~M}\binits{P.~M.}},
  \bauthor{\bsnm{Emonet},~\bfnm{Thierry}\binits{T.}} \AND
  \bauthor{\bsnm{Gerstein},~\bfnm{Mark~B}\binits{M.~B.}}
(\byear{2009}).
\btitle{Understanding modularity in molecular networks requires dynamics}.
\bjournal{Science signaling}
\bvolume{2}
\bpages{pe44}.
\end{barticle}
\endbibitem

\bibitem[\protect\citeauthoryear{Alsanie et~al.}{2017}]{alsanie2017homophilic}
\begin{barticle}[author]
\bauthor{\bsnm{Alsanie},~\bfnm{WF}\binits{W.}},
  \bauthor{\bsnm{Penna},~\bfnm{V}\binits{V.}},
  \bauthor{\bsnm{Schachner},~\bfnm{M}\binits{M.}},
  \bauthor{\bsnm{Thompson},~\bfnm{LH}\binits{L.}} \AND
  \bauthor{\bsnm{Parish},~\bfnm{CL}\binits{C.}}
(\byear{2017}).
\btitle{Homophilic binding of the neural cell adhesion molecule CHL1 regulates
  development of ventral midbrain dopaminergic pathways}.
\bjournal{Scientific reports}
\bvolume{7}
\bpages{9368}.
\end{barticle}
\endbibitem

\bibitem[\protect\citeauthoryear{Andrews and Mallows}{1974}]{andrews1974scale}
\begin{barticle}[author]
\bauthor{\bsnm{Andrews},~\bfnm{David~F}\binits{D.~F.}} \AND
  \bauthor{\bsnm{Mallows},~\bfnm{Colin~L}\binits{C.~L.}}
(\byear{1974}).
\btitle{Scale mixtures of normal distributions}.
\bjournal{Journal of the Royal Statistical Society. Series B (Methodological)}
\bvolume{36}
\bpages{99--102}.
\end{barticle}
\endbibitem

\bibitem[\protect\citeauthoryear{Carvalho, Polson and
  Scott}{2010}]{carvalho2010horseshoe}
\begin{barticle}[author]
\bauthor{\bsnm{Carvalho},~\bfnm{Carlos~M}\binits{C.~M.}},
  \bauthor{\bsnm{Polson},~\bfnm{Nicholas~G}\binits{N.~G.}} \AND
  \bauthor{\bsnm{Scott},~\bfnm{James~G}\binits{J.~G.}}
(\byear{2010}).
\btitle{The horseshoe estimator for sparse signals}.
\bjournal{Biometrika}
\bvolume{97}
\bpages{465--480}.
\end{barticle}
\endbibitem

\bibitem[\protect\citeauthoryear{Castillo et~al.}{2015}]{castillo2015bayesian}
\begin{barticle}[author]
\bauthor{\bsnm{Castillo},~\bfnm{Isma{\"e}l}\binits{I.}},
  \bauthor{\bsnm{Schmidt-Hieber},~\bfnm{Johannes}\binits{J.}},
  \bauthor{\bparticle{Van~der} \bsnm{Vaart},~\bfnm{Aad}\binits{A.}}
  \betal{et~al.}
(\byear{2015}).
\btitle{Bayesian linear regression with sparse priors}.
\bjournal{The Annals of Statistics}
\bvolume{43}
\bpages{1986--2018}.
\end{barticle}
\endbibitem

\bibitem[\protect\citeauthoryear{Cech and Steitz}{2014}]{cech2014noncoding}
\begin{barticle}[author]
\bauthor{\bsnm{Cech},~\bfnm{Thomas~R}\binits{T.~R.}} \AND
  \bauthor{\bsnm{Steitz},~\bfnm{Joan~A}\binits{J.~A.}}
(\byear{2014}).
\btitle{The noncoding RNA revolution—trashing old rules to forge new ones}.
\bjournal{Cell}
\bvolume{157}
\bpages{77--94}.
\end{barticle}
\endbibitem

\bibitem[\protect\citeauthoryear{Crane et~al.}{2016}]{crane2016dynamic}
\begin{barticle}[author]
\bauthor{\bsnm{Crane},~\bfnm{Harry}\binits{H.}} \betal{et~al.}
(\byear{2016}).
\btitle{Dynamic random networks and their graph limits}.
\bjournal{The Annals of Applied Probability}
\bvolume{26}
\bpages{691--721}.
\end{barticle}
\endbibitem

\bibitem[\protect\citeauthoryear{Durante et~al.}{2016}]{durante2016locally}
\begin{barticle}[author]
\bauthor{\bsnm{Durante},~\bfnm{Daniele}\binits{D.}},
  \bauthor{\bsnm{Dunson},~\bfnm{David~B}\binits{D.~B.}} \betal{et~al.}
(\byear{2016}).
\btitle{Locally adaptive dynamic networks}.
\bjournal{The Annals of Applied Statistics}
\bvolume{10}
\bpages{2203--2232}.
\end{barticle}
\endbibitem

\bibitem[\protect\citeauthoryear{Fan, Feng and Wu}{2009}]{fan2009network}
\begin{barticle}[author]
\bauthor{\bsnm{Fan},~\bfnm{Jianqing}\binits{J.}},
  \bauthor{\bsnm{Feng},~\bfnm{Yang}\binits{Y.}} \AND
  \bauthor{\bsnm{Wu},~\bfnm{Yichao}\binits{Y.}}
(\byear{2009}).
\btitle{Network exploration via the adaptive LASSO and SCAD penalties}.
\bjournal{The annals of applied statistics}
\bvolume{3}
\bpages{521}.
\end{barticle}
\endbibitem

\bibitem[\protect\citeauthoryear{Friedman, Hastie and
  Tibshirani}{2008}]{friedman2008sparse}
\begin{barticle}[author]
\bauthor{\bsnm{Friedman},~\bfnm{Jerome}\binits{J.}},
  \bauthor{\bsnm{Hastie},~\bfnm{Trevor}\binits{T.}} \AND
  \bauthor{\bsnm{Tibshirani},~\bfnm{Robert}\binits{R.}}
(\byear{2008}).
\btitle{Sparse inverse covariance estimation with the graphical lasso}.
\bjournal{Biostatistics}
\bvolume{9}
\bpages{432--441}.
\end{barticle}
\endbibitem

\bibitem[\protect\citeauthoryear{Geweke et~al.}{1991}]{geweke1991evaluating}
\begin{bbook}[author]
\bauthor{\bsnm{Geweke},~\bfnm{John}\binits{J.}} \betal{et~al.}
(\byear{1991}).
\btitle{Evaluating the accuracy of sampling-based approaches to the calculation
  of posterior moments}
\bvolume{196}.
\bpublisher{Federal Reserve Bank of Minneapolis, Research Department
  Minneapolis, MN}.
\end{bbook}
\endbibitem

\bibitem[\protect\citeauthoryear{Heidelberger and
  Welch}{1981}]{heidelberger1981spectral}
\begin{barticle}[author]
\bauthor{\bsnm{Heidelberger},~\bfnm{Philip}\binits{P.}} \AND
  \bauthor{\bsnm{Welch},~\bfnm{Peter~D}\binits{P.~D.}}
(\byear{1981}).
\btitle{A spectral method for confidence interval generation and run length
  control in simulations}.
\bjournal{Communications of the ACM}
\bvolume{24}
\bpages{233--245}.
\end{barticle}
\endbibitem

\bibitem[\protect\citeauthoryear{John et~al.}{2012}]{john2012bcl11a}
\begin{barticle}[author]
\bauthor{\bsnm{John},~\bfnm{Anita}\binits{A.}},
  \bauthor{\bsnm{Brylka},~\bfnm{Heike}\binits{H.}},
  \bauthor{\bsnm{Wiegreffe},~\bfnm{Christoph}\binits{C.}},
  \bauthor{\bsnm{Simon},~\bfnm{Ruth}\binits{R.}},
  \bauthor{\bsnm{Liu},~\bfnm{Pentao}\binits{P.}},
  \bauthor{\bsnm{J{\"u}ttner},~\bfnm{Ren{\'e}}\binits{R.}},
  \bauthor{\bsnm{Crenshaw},~\bfnm{E~Bryan}\binits{E.~B.}},
  \bauthor{\bsnm{Luyten},~\bfnm{Frank~P}\binits{F.~P.}},
  \bauthor{\bsnm{Jenkins},~\bfnm{Nancy~A}\binits{N.~A.}},
  \bauthor{\bsnm{Copeland},~\bfnm{Neal~G}\binits{N.~G.}} \betal{et~al.}
(\byear{2012}).
\btitle{Bcl11a is required for neuronal morphogenesis and sensory circuit
  formation in dorsal spinal cord development}.
\bjournal{Development}
\bvolume{139}
\bpages{1831--1841}.
\end{barticle}
\endbibitem

\bibitem[\protect\citeauthoryear{Kalaitzis
  et~al.}{2013}]{kalaitzis2013bigraphical}
\begin{binproceedings}[author]
\bauthor{\bsnm{Kalaitzis},~\bfnm{Alfredo}\binits{A.}},
  \bauthor{\bsnm{Lafferty},~\bfnm{John}\binits{J.}},
  \bauthor{\bsnm{Lawrence},~\bfnm{Neil}\binits{N.}} \AND
  \bauthor{\bsnm{Zhou},~\bfnm{Shuheng}\binits{S.}}
(\byear{2013}).
\btitle{The bigraphical lasso}.
In \bbooktitle{Proceedings of the 30th International Conference on Machine
  Learning (ICML-13)}
\bpages{1229--1237}.
\end{binproceedings}
\endbibitem

\bibitem[\protect\citeauthoryear{Kharchenko, Silberstein and
  Scadden}{2014}]{kharchenko2014bayesian}
\begin{barticle}[author]
\bauthor{\bsnm{Kharchenko},~\bfnm{Peter~V}\binits{P.~V.}},
  \bauthor{\bsnm{Silberstein},~\bfnm{Lev}\binits{L.}} \AND
  \bauthor{\bsnm{Scadden},~\bfnm{David~T}\binits{D.~T.}}
(\byear{2014}).
\btitle{Bayesian approach to single-cell differential expression analysis}.
\bjournal{Nature methods}
\bvolume{11}
\bpages{740--742}.
\end{barticle}
\endbibitem

\bibitem[\protect\citeauthoryear{Kolar et~al.}{2010}]{kolar2010estimating}
\begin{barticle}[author]
\bauthor{\bsnm{Kolar},~\bfnm{Mladen}\binits{M.}},
  \bauthor{\bsnm{Song},~\bfnm{Le}\binits{L.}},
  \bauthor{\bsnm{Ahmed},~\bfnm{Amr}\binits{A.}} \AND
  \bauthor{\bsnm{Xing},~\bfnm{Eric~P}\binits{E.~P.}}
(\byear{2010}).
\btitle{Estimating time-varying networks}.
\bjournal{The Annals of Applied Statistics}
\bvolume{4}
\bpages{94--123}.
\end{barticle}
\endbibitem

\bibitem[\protect\citeauthoryear{Kyung et~al.}{2010}]{kyung2010penalized}
\begin{barticle}[author]
\bauthor{\bsnm{Kyung},~\bfnm{Minjung}\binits{M.}},
  \bauthor{\bsnm{Gill},~\bfnm{Jeff}\binits{J.}},
  \bauthor{\bsnm{Ghosh},~\bfnm{Malay}\binits{M.}} \AND
  \bauthor{\bsnm{Casella},~\bfnm{George}\binits{G.}}
(\byear{2010}).
\btitle{Penalized regression, standard errors, and Bayesian lassos}.
\bjournal{Bayesian Analysis}
\bvolume{5}
\bpages{369--411}.
\end{barticle}
\endbibitem

\bibitem[\protect\citeauthoryear{Lauritzen}{1996}]{lauritzen1996graphical}
\begin{bbook}[author]
\bauthor{\bsnm{Lauritzen},~\bfnm{Steffen~L}\binits{S.~L.}}
(\byear{1996}).
\btitle{Graphical models}
\bvolume{17}.
\bpublisher{Clarendon Press}.
\end{bbook}
\endbibitem

\bibitem[\protect\citeauthoryear{Lebre et~al.}{2010}]{lebre2010statistical}
\begin{barticle}[author]
\bauthor{\bsnm{Lebre},~\bfnm{Sophie}\binits{S.}},
  \bauthor{\bsnm{Becq},~\bfnm{Jennifer}\binits{J.}},
  \bauthor{\bsnm{Devaux},~\bfnm{Frederic}\binits{F.}},
  \bauthor{\bsnm{Stumpf},~\bfnm{Michael~PH}\binits{M.~P.}} \AND
  \bauthor{\bsnm{Lelandais},~\bfnm{Gaelle}\binits{G.}}
(\byear{2010}).
\btitle{Statistical inference of the time-varying structure of gene-regulation
  networks}.
\bjournal{BMC systems biology}
\bvolume{4}
\bpages{130}.
\end{barticle}
\endbibitem

\bibitem[\protect\citeauthoryear{Li et~al.}{2008}]{li2008transcription}
\begin{barticle}[author]
\bauthor{\bsnm{Li},~\bfnm{Hao}\binits{H.}},
  \bauthor{\bsnm{Radford},~\bfnm{Jonathan~C}\binits{J.~C.}},
  \bauthor{\bsnm{Ragusa},~\bfnm{Michael~J}\binits{M.~J.}},
  \bauthor{\bsnm{Shea},~\bfnm{Katherine~L}\binits{K.~L.}},
  \bauthor{\bsnm{McKercher},~\bfnm{Scott~R}\binits{S.~R.}},
  \bauthor{\bsnm{Zaremba},~\bfnm{Jeffrey~D}\binits{J.~D.}},
  \bauthor{\bsnm{Soussou},~\bfnm{Walid}\binits{W.}},
  \bauthor{\bsnm{Nie},~\bfnm{Zhiguo}\binits{Z.}},
  \bauthor{\bsnm{Kang},~\bfnm{Yeon-Joo}\binits{Y.-J.}},
  \bauthor{\bsnm{Nakanishi},~\bfnm{Nobuki}\binits{N.}} \betal{et~al.}
(\byear{2008}).
\btitle{Transcription factor MEF2C influences neural stem/progenitor cell
  differentiation and maturation in vivo}.
\bjournal{Proceedings of the National Academy of Sciences}
\bvolume{105}
\bpages{9397--9402}.
\end{barticle}
\endbibitem

\bibitem[\protect\citeauthoryear{Linqing et~al.}{2015}]{linqing2015runx1t1}
\begin{barticle}[author]
\bauthor{\bsnm{Linqing},~\bfnm{Zou}\binits{Z.}},
  \bauthor{\bsnm{Guohua},~\bfnm{Jin}\binits{J.}},
  \bauthor{\bsnm{Haoming},~\bfnm{Li}\binits{L.}},
  \bauthor{\bsnm{Xuelei},~\bfnm{Tao}\binits{T.}},
  \bauthor{\bsnm{Jianbing},~\bfnm{Qin}\binits{Q.}} \AND
  \bauthor{\bsnm{Meiling},~\bfnm{Tian}\binits{T.}}
(\byear{2015}).
\btitle{RUNX1T1 regulates the neuronal differentiation of radial glial cells
  from the rat hippocampus}.
\bjournal{Stem cells translational medicine}
\bvolume{4}
\bpages{110--116}.
\end{barticle}
\endbibitem

\bibitem[\protect\citeauthoryear{Matias and
  Miele}{2016}]{matias2016statistical}
\begin{barticle}[author]
\bauthor{\bsnm{Matias},~\bfnm{Catherine}\binits{C.}} \AND
  \bauthor{\bsnm{Miele},~\bfnm{Vincent}\binits{V.}}
(\byear{2016}).
\btitle{Statistical clustering of temporal networks through a dynamic
  stochastic block model}.
\bjournal{Journal of the Royal Statistical Society: Series B (Statistical
  Methodology)}
\bvolume{79}
\bpages{1119-1141}.
\end{barticle}
\endbibitem

\bibitem[\protect\citeauthoryear{Matsumoto et~al.}{2017}]{matsumoto2017scode}
\begin{barticle}[author]
\bauthor{\bsnm{Matsumoto},~\bfnm{Hirotaka}\binits{H.}},
  \bauthor{\bsnm{Kiryu},~\bfnm{Hisanori}\binits{H.}},
  \bauthor{\bsnm{Furusawa},~\bfnm{Chikara}\binits{C.}},
  \bauthor{\bsnm{Ko},~\bfnm{Minoru~SH}\binits{M.~S.}},
  \bauthor{\bsnm{Ko},~\bfnm{Shigeru~BH}\binits{S.~B.}},
  \bauthor{\bsnm{Gouda},~\bfnm{Norio}\binits{N.}},
  \bauthor{\bsnm{Hayashi},~\bfnm{Tetsutaro}\binits{T.}} \AND
  \bauthor{\bsnm{Nikaido},~\bfnm{Itoshi}\binits{I.}}
(\byear{2017}).
\btitle{SCODE: an efficient regulatory network inference algorithm from
  single-cell RNA-Seq during differentiation}.
\bjournal{Bioinformatics}
\bvolume{33}
\bpages{2314--2321}.
\end{barticle}
\endbibitem

\bibitem[\protect\citeauthoryear{Mayer et~al.}{2019}]{mayer2019multimodal}
\begin{barticle}[author]
\bauthor{\bsnm{Mayer},~\bfnm{Simone}\binits{S.}},
  \bauthor{\bsnm{Chen},~\bfnm{Jiadong}\binits{J.}},
  \bauthor{\bsnm{Velmeshev},~\bfnm{Dmitry}\binits{D.}},
  \bauthor{\bsnm{Mayer},~\bfnm{Andreas}\binits{A.}},
  \bauthor{\bsnm{Eze},~\bfnm{Ugomma~C}\binits{U.~C.}},
  \bauthor{\bsnm{Bhaduri},~\bfnm{Aparna}\binits{A.}},
  \bauthor{\bsnm{Cunha},~\bfnm{Carlos~E}\binits{C.~E.}},
  \bauthor{\bsnm{Jung},~\bfnm{Diane}\binits{D.}},
  \bauthor{\bsnm{Arjun},~\bfnm{Arpana}\binits{A.}},
  \bauthor{\bsnm{Li},~\bfnm{Emmy}\binits{E.}} \betal{et~al.}
(\byear{2019}).
\btitle{Multimodal Single-Cell Analysis Reveals Physiological Maturation in the
  Developing Human Neocortex}.
\bjournal{Neuron}.
\end{barticle}
\endbibitem

\bibitem[\protect\citeauthoryear{Novershtern, Regev and
  Friedman}{2011}]{novershtern2011physical}
\begin{barticle}[author]
\bauthor{\bsnm{Novershtern},~\bfnm{Noa}\binits{N.}},
  \bauthor{\bsnm{Regev},~\bfnm{Aviv}\binits{A.}} \AND
  \bauthor{\bsnm{Friedman},~\bfnm{Nir}\binits{N.}}
(\byear{2011}).
\btitle{Physical Module Networks: an integrative approach for reconstructing
  transcription regulation}.
\bjournal{Bioinformatics}
\bvolume{27}
\bpages{i177--i185}.
\end{barticle}
\endbibitem

\bibitem[\protect\citeauthoryear{Nowakowski
  et~al.}{2017}]{nowakowski2017spatiotemporal}
\begin{barticle}[author]
\bauthor{\bsnm{Nowakowski},~\bfnm{Tomasz~J}\binits{T.~J.}},
  \bauthor{\bsnm{Bhaduri},~\bfnm{Aparna}\binits{A.}},
  \bauthor{\bsnm{Pollen},~\bfnm{Alex~A}\binits{A.~A.}},
  \bauthor{\bsnm{Alvarado},~\bfnm{Beatriz}\binits{B.}},
  \bauthor{\bsnm{Mostajo-Radji},~\bfnm{Mohammed~A}\binits{M.~A.}},
  \bauthor{\bsnm{Di~Lullo},~\bfnm{Elizabeth}\binits{E.}},
  \bauthor{\bsnm{Haeussler},~\bfnm{Maximilian}\binits{M.}},
  \bauthor{\bsnm{Sandoval-Espinosa},~\bfnm{Carmen}\binits{C.}},
  \bauthor{\bsnm{Liu},~\bfnm{Siyuan~John}\binits{S.~J.}},
  \bauthor{\bsnm{Velmeshev},~\bfnm{Dmitry}\binits{D.}},
  \bauthor{\bsnm{Ounadjela},~\bfnm{Johain~R}\binits{J.~R.}},
  \bauthor{\bsnm{Shuga},~\bfnm{Joe}\binits{J.}},
  \bauthor{\bsnm{Wang},~\bfnm{Xiaohui}\binits{X.}},
  \bauthor{\bsnm{Lim},~\bfnm{Daniel~A}\binits{D.~A.}},
  \bauthor{\bsnm{West},~\bfnm{Jay~A}\binits{J.~A.}},
  \bauthor{\bsnm{Leyrat},~\bfnm{Anne~A}\binits{A.~A.}},
  \bauthor{\bsnm{Kent},~\bfnm{W~James}\binits{W.~J.}} \AND
  \bauthor{\bsnm{Kriegstein},~\bfnm{Arnold~R}\binits{A.~R.}}
(\byear{2017}).
\btitle{Spatiotemporal gene expression trajectories reveal developmental
  hierarchies of the human cortex}.
\bjournal{Science}
\bvolume{358}
\bpages{1318--1323}.
\end{barticle}
\endbibitem

\bibitem[\protect\citeauthoryear{Palla, Caron and
  Teh}{2016}]{palla2016bayesian}
\begin{barticle}[author]
\bauthor{\bsnm{Palla},~\bfnm{Konstantina}\binits{K.}},
  \bauthor{\bsnm{Caron},~\bfnm{Francois}\binits{F.}} \AND
  \bauthor{\bsnm{Teh},~\bfnm{Yee~Whye}\binits{Y.~W.}}
(\byear{2016}).
\btitle{Bayesian nonparametrics for Sparse Dynamic Networks}.
\bjournal{arXiv preprint arXiv:1607.01624}.
\end{barticle}
\endbibitem

\bibitem[\protect\citeauthoryear{Park and Casella}{2008}]{park2008bayesian}
\begin{barticle}[author]
\bauthor{\bsnm{Park},~\bfnm{Trevor}\binits{T.}} \AND
  \bauthor{\bsnm{Casella},~\bfnm{George}\binits{G.}}
(\byear{2008}).
\btitle{The bayesian lasso}.
\bjournal{Journal of the American Statistical Association}
\bvolume{103}
\bpages{681--686}.
\end{barticle}
\endbibitem

\bibitem[\protect\citeauthoryear{Pataskar et~al.}{2016}]{pataskar2016neurod1}
\begin{barticle}[author]
\bauthor{\bsnm{Pataskar},~\bfnm{Abhijeet}\binits{A.}},
  \bauthor{\bsnm{Jung},~\bfnm{Johannes}\binits{J.}},
  \bauthor{\bsnm{Smialowski},~\bfnm{Pawel}\binits{P.}},
  \bauthor{\bsnm{Noack},~\bfnm{Florian}\binits{F.}},
  \bauthor{\bsnm{Calegari},~\bfnm{Federico}\binits{F.}},
  \bauthor{\bsnm{Straub},~\bfnm{Tobias}\binits{T.}} \AND
  \bauthor{\bsnm{Tiwari},~\bfnm{Vijay~K}\binits{V.~K.}}
(\byear{2016}).
\btitle{NeuroD1 reprograms chromatin and transcription factor landscapes to
  induce the neuronal program}.
\bjournal{The EMBO journal}
\bvolume{35}
\bpages{24--45}.
\end{barticle}
\endbibitem

\bibitem[\protect\citeauthoryear{Pensky}{2016}]{pensky2016dynamic}
\begin{barticle}[author]
\bauthor{\bsnm{Pensky},~\bfnm{Marianna}\binits{M.}}
(\byear{2016}).
\btitle{Dynamic network models and graphon estimation}.
\bjournal{arXiv preprint arXiv:1607.00673}.
\end{barticle}
\endbibitem

\bibitem[\protect\citeauthoryear{Pierson and Yau}{2015}]{pierson2015zifa}
\begin{barticle}[author]
\bauthor{\bsnm{Pierson},~\bfnm{Emma}\binits{E.}} \AND
  \bauthor{\bsnm{Yau},~\bfnm{Christopher}\binits{C.}}
(\byear{2015}).
\btitle{ZIFA: Dimensionality reduction for zero-inflated single-cell gene
  expression analysis}.
\bjournal{Genome biology}
\bvolume{16}
\bpages{1--10}.
\end{barticle}
\endbibitem

\bibitem[\protect\citeauthoryear{Piper et~al.}{2010}]{piper2010nfia}
\begin{barticle}[author]
\bauthor{\bsnm{Piper},~\bfnm{Michael}\binits{M.}},
  \bauthor{\bsnm{Barry},~\bfnm{Guy}\binits{G.}},
  \bauthor{\bsnm{Hawkins},~\bfnm{John}\binits{J.}},
  \bauthor{\bsnm{Mason},~\bfnm{Sharon}\binits{S.}},
  \bauthor{\bsnm{Lindwall},~\bfnm{Charlotta}\binits{C.}},
  \bauthor{\bsnm{Little},~\bfnm{Erica}\binits{E.}},
  \bauthor{\bsnm{Sarkar},~\bfnm{Anindita}\binits{A.}},
  \bauthor{\bsnm{Smith},~\bfnm{Aaron~G}\binits{A.~G.}},
  \bauthor{\bsnm{Moldrich},~\bfnm{Randal~X}\binits{R.~X.}},
  \bauthor{\bsnm{Boyle},~\bfnm{Glen~M}\binits{G.~M.}} \betal{et~al.}
(\byear{2010}).
\btitle{NFIA controls telencephalic progenitor cell differentiation through
  repression of the Notch effector Hes1}.
\bjournal{Journal of Neuroscience}
\bvolume{30}
\bpages{9127--9139}.
\end{barticle}
\endbibitem

\bibitem[\protect\citeauthoryear{Plummer et~al.}{2006}]{plum2006coda}
\begin{barticle}[author]
\bauthor{\bsnm{Plummer},~\bfnm{Martyn}\binits{M.}},
  \bauthor{\bsnm{Best},~\bfnm{Nicky}\binits{N.}},
  \bauthor{\bsnm{Cowles},~\bfnm{Kate}\binits{K.}} \AND
  \bauthor{\bsnm{Vines},~\bfnm{Karen}\binits{K.}}
(\byear{2006}).
\btitle{CODA: Convergence Diagnosis and Output Analysis for MCMC}.
\bjournal{R News}
\bvolume{6}
\bpages{7--11}.
\end{barticle}
\endbibitem

\bibitem[\protect\citeauthoryear{Qiu et~al.}{2011}]{qiu2011extracting}
\begin{barticle}[author]
\bauthor{\bsnm{Qiu},~\bfnm{Peng}\binits{P.}},
  \bauthor{\bsnm{Simonds},~\bfnm{Erin~F}\binits{E.~F.}},
  \bauthor{\bsnm{Bendall},~\bfnm{Sean~C}\binits{S.~C.}},
  \bauthor{\bsnm{Gibbs~Jr},~\bfnm{Kenneth~D}\binits{K.~D.}},
  \bauthor{\bsnm{Bruggner},~\bfnm{Robert~V}\binits{R.~V.}},
  \bauthor{\bsnm{Linderman},~\bfnm{Michael~D}\binits{M.~D.}},
  \bauthor{\bsnm{Sachs},~\bfnm{Karen}\binits{K.}},
  \bauthor{\bsnm{Nolan},~\bfnm{Garry~P}\binits{G.~P.}} \AND
  \bauthor{\bsnm{Plevritis},~\bfnm{Sylvia~K}\binits{S.~K.}}
(\byear{2011}).
\btitle{Extracting a cellular hierarchy from high-dimensional cytometry data
  with SPADE}.
\bjournal{Nature biotechnology}
\bvolume{29}
\bpages{886--891}.
\end{barticle}
\endbibitem

\bibitem[\protect\citeauthoryear{Rehfeld et~al.}{2018}]{rehfeld2018rna}
\begin{barticle}[author]
\bauthor{\bsnm{Rehfeld},~\bfnm{Frederick}\binits{F.}},
  \bauthor{\bsnm{Maticzka},~\bfnm{Daniel}\binits{D.}},
  \bauthor{\bsnm{Grosser},~\bfnm{Sabine}\binits{S.}},
  \bauthor{\bsnm{Knauff},~\bfnm{Pina}\binits{P.}},
  \bauthor{\bsnm{Eravci},~\bfnm{Murat}\binits{M.}},
  \bauthor{\bsnm{Vida},~\bfnm{Imre}\binits{I.}},
  \bauthor{\bsnm{Backofen},~\bfnm{Rolf}\binits{R.}} \AND
  \bauthor{\bsnm{Wulczyn},~\bfnm{F~Gregory}\binits{F.~G.}}
(\byear{2018}).
\btitle{The RNA-binding protein ARPP21 controls dendritic branching by
  functionally opposing the miRNA it hosts}.
\bjournal{Nature communications}
\bvolume{9}
\bpages{1235}.
\end{barticle}
\endbibitem

\bibitem[\protect\citeauthoryear{Rosengren and
  Trapman}{2016}]{rosengren2016dynamic}
\begin{barticle}[author]
\bauthor{\bsnm{Rosengren},~\bfnm{Sebastian}\binits{S.}} \AND
  \bauthor{\bsnm{Trapman},~\bfnm{Pieter}\binits{P.}}
(\byear{2016}).
\btitle{A Dynamic Erdos Renyi Graph Model}.
\bjournal{arXiv preprint arXiv:1604.05127}.
\end{barticle}
\endbibitem

\bibitem[\protect\citeauthoryear{Sarkar and
  Chakrabarti}{2014}]{sarkar2014nonparametric}
\begin{barticle}[author]
\bauthor{\bsnm{Sarkar},~\bfnm{Purnamrita}\binits{P.}} \AND
  \bauthor{\bsnm{Chakrabarti},~\bfnm{Deepayan}\binits{D.}}
(\byear{2014}).
\btitle{Nonparametric link prediction in large scale dynamic networks}.
\bjournal{Electronic Journal of Statistics}
\bvolume{8}
\bpages{2022--2065}.
\end{barticle}
\endbibitem

\bibitem[\protect\citeauthoryear{Schaefer et~al.}{2014}]{schaefer2014dynamic}
\begin{barticle}[author]
\bauthor{\bsnm{Schaefer},~\bfnm{Alexander}\binits{A.}},
  \bauthor{\bsnm{Margulies},~\bfnm{Daniel~S}\binits{D.~S.}},
  \bauthor{\bsnm{Lohmann},~\bfnm{Gabriele}\binits{G.}},
  \bauthor{\bsnm{Gorgolewski},~\bfnm{Krzysztof~J}\binits{K.~J.}},
  \bauthor{\bsnm{Smallwood},~\bfnm{Jonathan}\binits{J.}},
  \bauthor{\bsnm{Kiebel},~\bfnm{Stefan~J}\binits{S.~J.}} \AND
  \bauthor{\bsnm{Villringer},~\bfnm{Arno}\binits{A.}}
(\byear{2014}).
\btitle{Dynamic network participation of functional connectivity hubs assessed
  by resting-state fMRI}.
\bjournal{Frontiers in human neuroscience}
\bvolume{8}
\bpages{195}.
\end{barticle}
\endbibitem

\bibitem[\protect\citeauthoryear{Sekara, Stopczynski and
  Lehmann}{2016}]{sekara2016fundamental}
\begin{barticle}[author]
\bauthor{\bsnm{Sekara},~\bfnm{Vedran}\binits{V.}},
  \bauthor{\bsnm{Stopczynski},~\bfnm{Arkadiusz}\binits{A.}} \AND
  \bauthor{\bsnm{Lehmann},~\bfnm{Sune}\binits{S.}}
(\byear{2016}).
\btitle{Fundamental structures of dynamic social networks}.
\bjournal{Proceedings of the national academy of sciences}
\bvolume{113}
\bpages{9977--9982}.
\end{barticle}
\endbibitem

\bibitem[\protect\citeauthoryear{Shimamura
  et~al.}{2016}]{shimamura2016bayesian}
\begin{barticle}[author]
\bauthor{\bsnm{Shimamura},~\bfnm{Kaito}\binits{K.}},
  \bauthor{\bsnm{Ueki},~\bfnm{Masao}\binits{M.}},
  \bauthor{\bsnm{Kawano},~\bfnm{Shuichi}\binits{S.}} \AND
  \bauthor{\bsnm{Konishi},~\bfnm{Sadanori}\binits{S.}}
(\byear{2016}).
\btitle{Bayesian generalized fused lasso modeling via NEG distribution}.
\bjournal{arXiv preprint arXiv:1602.04910}.
\end{barticle}
\endbibitem

\bibitem[\protect\citeauthoryear{Suv{\`a}
  et~al.}{2014}]{suva2014reconstructing}
\begin{barticle}[author]
\bauthor{\bsnm{Suv{\`a}},~\bfnm{Mario~L}\binits{M.~L.}},
  \bauthor{\bsnm{Rheinbay},~\bfnm{Esther}\binits{E.}},
  \bauthor{\bsnm{Gillespie},~\bfnm{Shawn~M}\binits{S.~M.}},
  \bauthor{\bsnm{Patel},~\bfnm{Anoop~P}\binits{A.~P.}},
  \bauthor{\bsnm{Wakimoto},~\bfnm{Hiroaki}\binits{H.}},
  \bauthor{\bsnm{Rabkin},~\bfnm{Samuel~D}\binits{S.~D.}},
  \bauthor{\bsnm{Riggi},~\bfnm{Nicolo}\binits{N.}},
  \bauthor{\bsnm{Chi},~\bfnm{Andrew~S}\binits{A.~S.}},
  \bauthor{\bsnm{Cahill},~\bfnm{Daniel~P}\binits{D.~P.}},
  \bauthor{\bsnm{Nahed},~\bfnm{Brian~V}\binits{B.~V.}} \betal{et~al.}
(\byear{2014}).
\btitle{Reconstructing and reprogramming the tumor-propagating potential of
  glioblastoma stem-like cells}.
\bjournal{Cell}
\bvolume{157}
\bpages{580--594}.
\end{barticle}
\endbibitem

\bibitem[\protect\citeauthoryear{Tibshirani
  et~al.}{2005}]{tibshirani2005sparsity}
\begin{barticle}[author]
\bauthor{\bsnm{Tibshirani},~\bfnm{Robert}\binits{R.}},
  \bauthor{\bsnm{Saunders},~\bfnm{Michael}\binits{M.}},
  \bauthor{\bsnm{Rosset},~\bfnm{Saharon}\binits{S.}},
  \bauthor{\bsnm{Zhu},~\bfnm{Ji}\binits{J.}} \AND
  \bauthor{\bsnm{Knight},~\bfnm{Keith}\binits{K.}}
(\byear{2005}).
\btitle{Sparsity and smoothness via the fused lasso}.
\bjournal{Journal of the Royal Statistical Society: Series B (Statistical
  Methodology)}
\bvolume{67}
\bpages{91--108}.
\end{barticle}
\endbibitem

\bibitem[\protect\citeauthoryear{Tibshirani
  et~al.}{2011}]{tibshirani2011solution}
\begin{bbook}[author]
\bauthor{\bsnm{Tibshirani},~\bfnm{Ryan~Joseph}\binits{R.~J.}},
  \bauthor{\bsnm{Taylor},~\bfnm{Jonathan~E}\binits{J.~E.}},
  \bauthor{\bsnm{Candes},~\bfnm{Emmanuel~Jean}\binits{E.~J.}} \AND
  \bauthor{\bsnm{Hastie},~\bfnm{Trevor}\binits{T.}}
(\byear{2011}).
\btitle{The solution path of the generalized lasso}.
\bpublisher{Stanford University}.
\end{bbook}
\endbibitem

\bibitem[\protect\citeauthoryear{Toriyama et~al.}{2006}]{toriyama2006shootin1}
\begin{barticle}[author]
\bauthor{\bsnm{Toriyama},~\bfnm{Michinori}\binits{M.}},
  \bauthor{\bsnm{Shimada},~\bfnm{Tadayuki}\binits{T.}},
  \bauthor{\bsnm{Kim},~\bfnm{Ki~Bum}\binits{K.~B.}},
  \bauthor{\bsnm{Mitsuba},~\bfnm{Mari}\binits{M.}},
  \bauthor{\bsnm{Nomura},~\bfnm{Eiko}\binits{E.}},
  \bauthor{\bsnm{Katsuta},~\bfnm{Kazuhiro}\binits{K.}},
  \bauthor{\bsnm{Sakumura},~\bfnm{Yuichi}\binits{Y.}},
  \bauthor{\bsnm{Roepstorff},~\bfnm{Peter}\binits{P.}} \AND
  \bauthor{\bsnm{Inagaki},~\bfnm{Naoyuki}\binits{N.}}
(\byear{2006}).
\btitle{Shootin1: A protein involved in the organization of an asymmetric
  signal for neuronal polarization}.
\bjournal{The Journal of cell biology}
\bvolume{175}
\bpages{147--157}.
\end{barticle}
\endbibitem

\bibitem[\protect\citeauthoryear{Trapnell et~al.}{2014}]{trapnell2014dynamics}
\begin{barticle}[author]
\bauthor{\bsnm{Trapnell},~\bfnm{Cole}\binits{C.}},
  \bauthor{\bsnm{Cacchiarelli},~\bfnm{Davide}\binits{D.}},
  \bauthor{\bsnm{Grimsby},~\bfnm{Jonna}\binits{J.}},
  \bauthor{\bsnm{Pokharel},~\bfnm{Prapti}\binits{P.}},
  \bauthor{\bsnm{Li},~\bfnm{Shuqiang}\binits{S.}},
  \bauthor{\bsnm{Morse},~\bfnm{Michael}\binits{M.}},
  \bauthor{\bsnm{Lennon},~\bfnm{Niall~J}\binits{N.~J.}},
  \bauthor{\bsnm{Livak},~\bfnm{Kenneth~J}\binits{K.~J.}},
  \bauthor{\bsnm{Mikkelsen},~\bfnm{Tarjei~S}\binits{T.~S.}} \AND
  \bauthor{\bsnm{Rinn},~\bfnm{John~L}\binits{J.~L.}}
(\byear{2014}).
\btitle{The dynamics and regulators of cell fate decisions are revealed by
  pseudotemporal ordering of single cells}.
\bjournal{Nature biotechnology}
\bvolume{32}
\bpages{381--386}.
\end{barticle}
\endbibitem

\bibitem[\protect\citeauthoryear{van~der Pas et~al.}{2016}]{van2016conditions}
\begin{barticle}[author]
\bauthor{\bparticle{van~der} \bsnm{Pas},~\bfnm{SL}\binits{S.}},
  \bauthor{\bsnm{Salomond},~\bfnm{J-B}\binits{J.-B.}},
  \bauthor{\bsnm{Schmidt-Hieber},~\bfnm{Johannes}\binits{J.}} \betal{et~al.}
(\byear{2016}).
\btitle{Conditions for posterior contraction in the sparse normal means
  problem}.
\bjournal{Electronic journal of statistics}
\bvolume{10}
\bpages{976--1000}.
\end{barticle}
\endbibitem

\bibitem[\protect\citeauthoryear{van Dijk et~al.}{2017}]{van2017magic}
\begin{barticle}[author]
\bauthor{\bparticle{van} \bsnm{Dijk},~\bfnm{David}\binits{D.}},
  \bauthor{\bsnm{Nainys},~\bfnm{Juozas}\binits{J.}},
  \bauthor{\bsnm{Sharma},~\bfnm{Roshan}\binits{R.}},
  \bauthor{\bsnm{Kathail},~\bfnm{Pooja}\binits{P.}},
  \bauthor{\bsnm{Carr},~\bfnm{Ambrose~J}\binits{A.~J.}},
  \bauthor{\bsnm{Moon},~\bfnm{Kevin~R}\binits{K.~R.}},
  \bauthor{\bsnm{Mazutis},~\bfnm{Linas}\binits{L.}},
  \bauthor{\bsnm{Wolf},~\bfnm{Guy}\binits{G.}},
  \bauthor{\bsnm{Krishnaswamy},~\bfnm{Smita}\binits{S.}} \AND
  \bauthor{\bsnm{Pe'er},~\bfnm{Dana}\binits{D.}}
(\byear{2017}).
\btitle{MAGIC: A diffusion-based imputation method reveals gene-gene
  interactions in single-cell RNA-sequencing data}.
\bjournal{BioRxiv}
\bpages{111591}.
\end{barticle}
\endbibitem

\bibitem[\protect\citeauthoryear{Xu and Hero~III}{2013}]{xu2013dynamic}
\begin{binproceedings}[author]
\bauthor{\bsnm{Xu},~\bfnm{Kevin~S}\binits{K.~S.}} \AND
  \bauthor{\bsnm{Hero~III},~\bfnm{Alfred~O}\binits{A.~O.}}
(\byear{2013}).
\btitle{Dynamic stochastic blockmodels: Statistical models for time-evolving
  networks}.
In \bbooktitle{International Conference on Social Computing,
  Behavioral-Cultural Modeling, and Prediction}
\bpages{201--210}.
\bpublisher{Springer}.
\end{binproceedings}
\endbibitem

\bibitem[\protect\citeauthoryear{Zhang, Zhao and Zhang}{2012}]{zhang2012common}
\begin{barticle}[author]
\bauthor{\bsnm{Zhang},~\bfnm{Shihua}\binits{S.}},
  \bauthor{\bsnm{Zhao},~\bfnm{Junfei}\binits{J.}} \AND
  \bauthor{\bsnm{Zhang},~\bfnm{Xiang-Sun}\binits{X.-S.}}
(\byear{2012}).
\btitle{Common community structure in time-varying networks}.
\bjournal{Physical Review E}
\bvolume{85}
\bpages{056110}.
\end{barticle}
\endbibitem

\end{thebibliography}

\clearpage
\includepdf[pages=-]{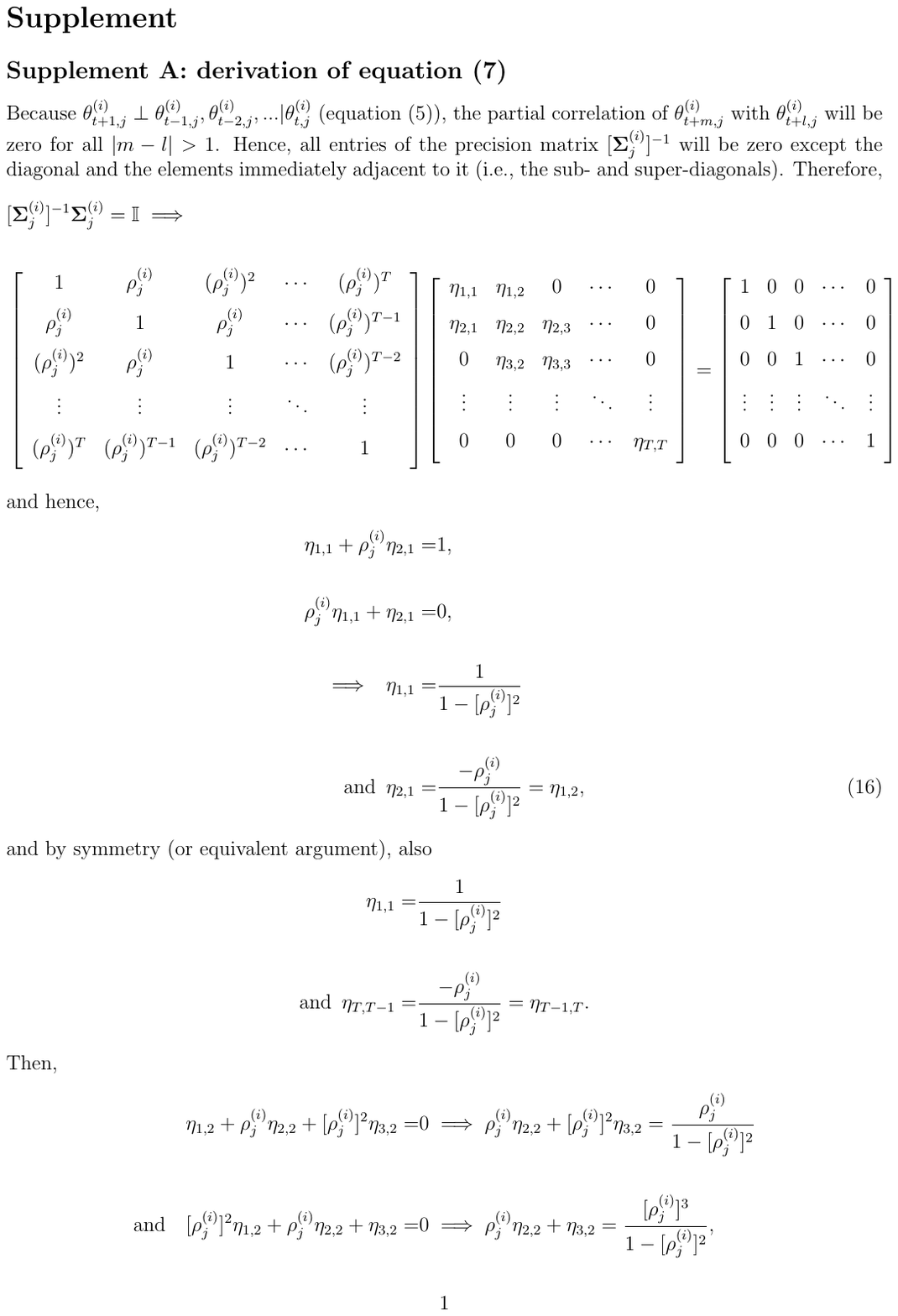}

\end{document}